%% file: ms.tex
\pdfoutput=1 

\documentclass[a4paper, 11pt, final]{article}

\input{ms_a_preamble}

\begin{document}


\renewcommand{\figureautorefname}{Fig.}
\onehalfspacing


\input{ms_b_frontmatter}



\newpage

\input{1-Intro}
\input{2-Method}
\input{3-tROC}

\input{4-DataCalibration}

\input{5-Results}

\input{6-Conclusion}


\appendix
\input{7-Appendix}


\singlespacing
\printbibliography 
\onehalfspacing



\end{document}

%% file: ms_a_preamble.tex
\usepackage{amssymb} 
\usepackage{amsthm} 
\usepackage{amsmath,empheq}
\usepackage{bbm}

\DeclareMathAlphabet{\mathcal}{OMS}{cmsy}{m}{n}
\DeclareMathAlphabet\mathbfcal{OMS}{cmsy}{b}{n}

\DeclareFontFamily{U}{dutchcal}{\skewchar\font=45 }
\DeclareFontShape{U}{dutchcal}{m}{n}{<-> s*[1.0] dutchcal-r}{}
\DeclareFontShape{U}{dutchcal}{b}{n}{<-> s*[1.0] dutchcal-b}{}
\DeclareMathAlphabet{\mathcald}{U}{dutchcal}{m}{n}
\SetMathAlphabet{\mathcald}{bold}{U}{dutchcal}{b}{n}
\DeclareMathAlphabet\mathcalz{T1}{pzc}{mb}{it}

\usepackage[nice]{nicefrac} 
\usepackage{siunitx} 

\usepackage{newtxtext, newtxmath} 

\usepackage{anyfontsize}
\usepackage[utf8]{inputenc}
\usepackage[T1]{fontenc}
\usepackage{microtype} 

\providecommand{\JEL}[1]{\textit{\textbf{JEL: }} #1}
\providecommand{\keywords}[1]{\textbf{\textit{Keywords--- }} #1}

\usepackage{setspace} 

\usepackage{parskip} 
\usepackage{etoolbox}
\usepackage{titlesec}
\titleformat{\section}{\normalfont\Large\bfseries}{\thesection}{1em}{}
\titleformat{\subsection}{\normalfont\large\bfseries}{\thesubsection.}{1em}{}
\titleformat{\subsubsection}{\normalfont\normalsize\itshape}{\thesubsubsection.}{1em}{}

\usepackage{abstract}

\renewenvironment{abstract}
 {\normalfont
  \begin{center}
  \bfseries \abstractname\vspace{-.5em}\vspace{0pt}
  \end{center}
  \list{}{
    \setlength{\leftmargin}{0cm}%
    \setlength{\rightmargin}{\leftmargin}%
  }%
  \item\relax}
 {\endlist}

\usepackage{authblk} 
\usepackage[bottom]{footmisc} 

\usepackage{multicol} 

\usepackage{enumitem} 

\usepackage{graphicx} 
\usepackage{float} 
\usepackage{subcaption}
\usepackage{afterpage} 

\usepackage{printlen}

\usepackage[labelfont=bf]{caption}
\usepackage{color, colortbl}
\definecolor{LightGray}{rgb}{0.93,0.914,0.914}    
\usepackage{soul} 
\usepackage{longtable,rotating} 
\usepackage{booktabs} 
\usepackage{multirow} 
\usepackage{arydshln} 
\usepackage{bigdelim} 

\usepackage[most]{tcolorbox}            

\PassOptionsToPackage{hyphens}{url} 

\usepackage{fancyhdr}
\fancyheadoffset{0cm}
\makeatletter
\newcommand{\quickwordcount}[1]{
  \immediate\write18{texcount -quiet -incbib -sub=none -utf8 -1 -sum -merge -encoding=utf8 #1.tex > #1-words}%
  \immediate\openin\somefile=#1-words
  \read\somefile to \@@localdummy
  \immediate\closein\somefile
  \setcounter{wordcounter}{\@@localdummy}
  \@@localdummy
}
\makeatother

\usepackage{silence}
\usepackage[style=apa,backend=biber,natbib,hyperref]{biblatex}
\DeclareLanguageMapping{british}{british-apa}
\defbibenvironment{bibliography}{\enumerate}{\endenumerate}{\item}

\usepackage[colorlinks=true,allcolors=gray]{hyperref} 
\let\orgautoref\autoref

\renewcommand{\autoref}[1]
{%
\def\equationautorefname{Eq.}%
\def\sectionautorefname{Sec.}%
\def\subsectionautorefname{Subsec.}%
\def\figureautorefname{Fig.}%
\def\subfigureautorefname{Fig.}%
\orgautoref{#1}%
}

\usepackage[nameinlink,capitalise]{cleveref}

\usepackage{algorithm,algpseudocode}

\makeatletter
\newlength{\trianglerightwidth}
\algnewcommand{\LineCommentCont}[1]{\Statex \hskip\ALG@thistlm%
  \parbox[t]{\dimexpr\linewidth-\ALG@thistlm}
{\leftskip=\algorithmicindent
  \hangindent=\algorithmicindent 
  \hangafter=1%
  \strut\makebox[\algorithmicindent][c]{$\triangleright$}#1\strut}
  } 
\makeatother

\usepackage[left=1.8cm, right=1.8cm, bottom=2.5cm, top=2.5cm]{geometry}

\usepackage{listings}
\usepackage{xcolor}

%% file: ms_b_frontmatter.tex

\newcommand{\MainTitleText}{
Exploring different subtypes of recurrent event Cox-regression models in modelling lifetime default risk: A tutorial
}

\title{\fontsize{20pt}{0pt}\selectfont\textbf{\MainTitleText
}}


\author[,a]{\large Arno Botha \thanks{ ORC iD: 0000-0002-1708-0153; Corresponding author: \url{arno.spasie.botha@gmail.com}}}
\author[,a,b]{\large Tanja Verster \thanks{ ORC iD: 0000-0002-4711-6145; email: \url{tanja.verster@nwu.ac.za}}}
\author[,a]{\large Bernard Scheepers \thanks{ ORC iD: 0009-0009-3670-5073}}
\affil[a]{\footnotesize \textit{Centre for Business Mathematics and Informatics \& Unit for Data Science and Computing, North-West University, Potchefstroom, South Africa}}
\affil[b]{\footnotesize \textit{National Institute for Theoretical and Computational Sciences (NITheCS), Potchefstroom, South Africa}}
\renewcommand\Authands{, and }

    

\makeatletter
\renewcommand{\@maketitle}{
    \newpage
     \null
     \vskip 1em%
     \begin{center}%
      {\LARGE \@title \par
      	\@author \par
        }
     \end{center}%
     \par
 } 
 \makeatother
 
 \maketitle

{
    \setlength{\parindent}{0cm}
    \rule{1\columnwidth}{0.4pt}
    \begin{abstract}
    In the pursuit of modelling a loan's probability of default (PD) over its lifetime, repeat default events are often ignored when using Cox Proportional Hazard (PH) models. Excluding such events may produce biased and inaccurate PD-estimates, which can compromise financial buffers against future losses. Accordingly, we investigate a few subtypes of Cox-models that can incorporate recurrent default events. We explore both the Andersen-Gill (AG) and the Prentice-Williams-Peterson (PWP) spell-time models using real-world data as an illustration. These models are compared against a baseline that deliberately ignores recurrent events, called the time to first default (TFD) model. Our models are evaluated using Harrell's c-statistic, adjusted Cox-Sell residuals, and a novel extension of time-dependent receiver operating characteristic analysis. From these Cox-models, we demonstrate how to derive a portfolio-level term-structure of default risk, which is a series of marginal PD-estimates over the average loan's lifetime. While the TFD-- and PWP-models do not differ significantly across all diagnostics, the AG-model underperformed expectations. We believe that our pedagogical tutorial, as accompanied by a codebase, would be of great value to practitioner and regulator alike. Accordingly, our work enhances the current practice of using Cox-modelling in producing timeous and accurate PD-estimates under IFRS 9.
    \end{abstract}
     
    \keywords{Credit risk; IFRS 9; Lifetime probability of default; Survival analysis; Recurrent events.}
     
     \JEL{C33, C41, C52, G21.}
    
    \rule{1\columnwidth}{0.4pt}
}

\noindent Word count (excluding front matter and appendices): 9162 

\subsection*{Disclosure of interest and declaration of funding}
\noindent This work is financially supported wholly/in part by the National Research Foundation of South Africa (Grant Number 126885), with no known conflicts of interest that may have influenced the outcome of this work. The authors would like to thank all anonymous referees and editors for their extremely valuable contributions that have substantially improved this work.

%% file: 1-Intro.tex
\section{Introduction and literature review}
\label{sec:intro}

In granting loans, a bank faces the principal risk of losing the lent capital if the borrower fails to repay the loan. Predicting the \textit{probability of default} (PD) accurately is therefore paramount to the many decision-making processes within a bank. This prediction task involves the modelling of past defaults as a function of a set of borrower-specific input variables, which is a widely-studied problem; see \citet{hand1997statistical}, \citet{siddiqi2005credit}, \citet{thomas2009consumer}, \citet{hao2010review},  \citet{baesens2016credit}, and \citet{louzada2016review}.
However, \citet{crook2010dynamic} and \citet{botha2025multistate} argued that these credit rating systems generally produce a rather static PD-estimate that does not vary much over the lifetime of each loan or the economic cycle. This design flaw is however deliberate in complying with the Basel framework of the \citet{basel2019}, since its goal is to estimate stable levels of capital to absorb catastrophic (or unexpected) losses. That said, these stable PD-estimates are typically inaccurate and do not agree vociferously with reality, which renders them inappropriate within any other context besides capital estimation. Other contexts might very well warrant embedding any temporal effects that can affect the PD during loan life, together with those of the broader macroeconomic environment. In the interest of brevity, one may group these contexts across the typical 5-phase credit life cycle: marketing, acquisition, customer management, collection, and debt recovery; see \citet[\S 3.1.2]{botha2021phd}. Within each phase, various modelling exercises may either directly embed PD-estimates such as risk-based pricing, or rely indirectly on these PD-estimates such as scoring the collection success of defaulted loans. Inaccurate PD-estimates may therefore propagate and compromise any downstream decision-making system that may use such estimates. 
Accordingly, the ubiquitous utility of accurate and more dynamic PD-modelling is the premise on which we shall build this study.

The impetus for more dynamic PD-modelling was furthered by the introduction of IFRS 9 by the \citet{ifrs9_2014}. This accounting standard requires a financial asset's value to be comprehensively adjusted over its lifetime and in line with the bank's time-dependent expectation of the asset's \textit{credit risk}, i.e., the potential loss that may be induced by defaulting loans. As such, a bank forfeits a portion of its income into a centralised loss provision that can ideally offset the future write-off of troubled loans. This provision's size is regularly updated based on a statistical model of the loan's \textit{expected credit loss} (ECL), which typically embeds the PD as a risk parameter. Based on the ECL-estimate, a bank adjusts its provision either by forfeiting more income or by releasing a portion thereof back into the income statement; see \S 5.5.8 in IFRS 9. In calculating this ECL-estimate, one follows a 3-stage approach (\S 5.5.3 and \S 5.5.5) that is based on the extent of the perceived deterioration (or improvement) in credit risk. Principally, the ECL-estimate should become progressively more severe as a loan transits across these stages. In so doing, the recognition of credit losses should itself become more timeous and dynamic, which is the main imperative of IFRS 9; see \citet{PwC_2014}, \citet{EY_2018}, and \citet{botha2025sicr}.
Inaccurate PD-estimates would therefore directly affect a bank's income statement via the ECL-model, which can detract from the very spirit of IFRS 9.

According to \citet{skoglund2017}, a risk model can achieve such dynamicity in producing PD-estimates when it can project default risk across various time horizons over a loan's lifetime $\mathcal{T}$, and in tandem with forward-looking macroeconomic changes. This projection requires the estimation of a series of marginal PD-values at each discrete loan period $t=t_1,\dots,\mathcal{T}$, starting from the loan's time of initial recognition $t_1$. In turn, each marginal PD-estimate originates from a model as a function of a rich set of input variables, including macroeconomic covariates. 
As surveyed by \citet{crook2010dynamic}, there exists a small suite of loan-level modelling techniques that can produce such a series of PD-estimates over loan life. We shall call any such a series the \textit{term-structure} of default risk, which is typically a non-linear and right-skewed function of loan age, as demonstrated later. Put differently, default risk typically rises drastically over earlier times, whereafter it gradually dissipates again over time. 
This non-linearity testifies to the dynamicity of default risk as time progresses, which may itself shift in line with material macroeconomic events or loan-specific situations. One might even argue that IFRS 9 indirectly requires such non-linearity in producing ECL-estimates that are unbiased, time-dependent, and forward-looking; see \S 5.5.17 in IFRS 9.

However, the modelling of such a dynamic and time-dependent collection of PD-estimates is fraught with challenges. Perhaps the greatest challenge is due to the fact that `default' is not necessarily an absorbing state into which a loan is forever trapped, as discussed by \citet[pp.~73-83]{botha2021phd}. In fact, a loan may exit default (a phenomenon known as `curing') and be subject to default risk again, during which time it can default once more; a cycle that can repeat multiple times. 
This dynamicity is recognised in both \S36.74 of the Basel framework and in Article 178(5) of the Capital Requirements Regulation (CRR), as promulgated by the \citet{eu2013CRR} for the EU-market. Both pieces of legislation require a bank to rate loans as performing whenever default criteria cease to apply. 
This requirement therefore references the various cycles of curing and re-defaulting over a loan's lifetime (where applicable), which implies that the underlying credit risk models should ideally cater for this cyclic aspect in producing suitably dynamic PD-estimates.

One particularly powerful class of modelling techniques for rendering such dynamic and time-dependent PD-estimates is that of \textit{survival analysis}. By examining the length of time until reaching some well-defined endpoint (such as default), survival models can predict both the occurrence and timing of the main event; see \citet{singer1993time}, \citet{kleinbaum2012survival}, \citet{kartsonaki2016survival}, and \citet{schober2018survival} for an overview. 
While typically used in the biostatistical literature, \citet{narain1992credit} first modelled the PD using a survival model. \citet{banasik1999not} expanded thereon by estimating the PD as a function of input variables via a Cox proportional hazards regression model, which compared favourably to a logistic regression (LR) model. \citet{stepanova2002survival} further investigated certain modelling practices and associated diagnostics (e.g., Cox-Snell and Schoenfeld residuals) when using a Cox PH-model for PD-estimation. Other authors have shown that a Cox regression model for PD-estimation can be further improved by including time-dependent variables, especially macroeconomic ones; see \citet{bellotti2009macro}, \citet{crook2010dynamic}, \citet{bellotti2013}, and \citet{bellotti2014stresstesting}. Finally, \citet{dirick2017time} benchmarked a few survival model subtypes, including Cox regression with/without spline functions, accelerated failure time models, and mixture cure models.
This growing body of literature bodes well for further exploring the use of survival analysis in PD-estimation.

\begin{figure}[ht!]
    \centering
    \includegraphics[width=1\linewidth, height=0.38\textheight]{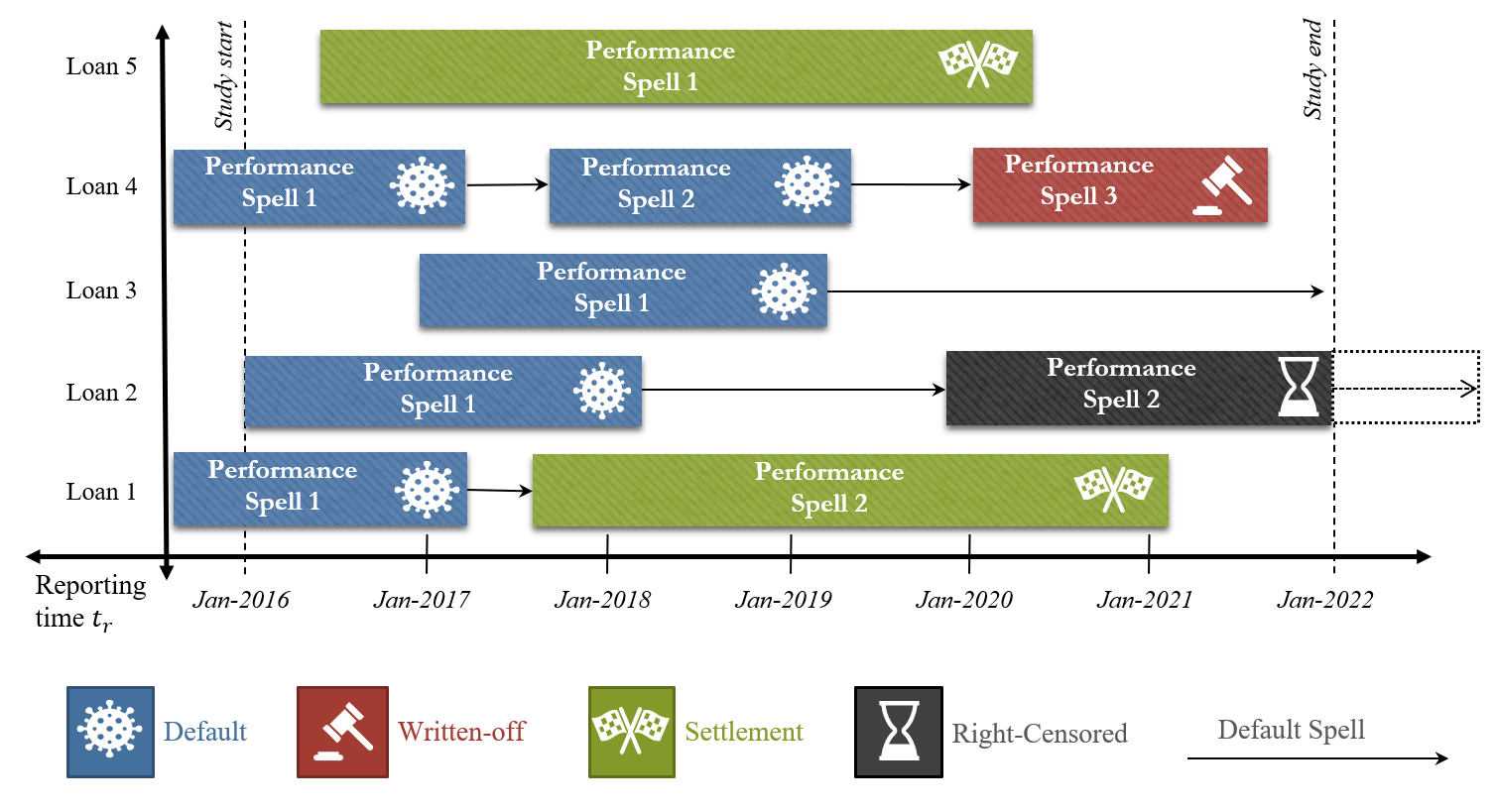}
    \caption{Demonstrating the resolution types and recurrence of performing spells over time for a few hypothetical loans.}
    \label{fig:PerfSpells}
\end{figure}

Despite their utility, the aforementioned studies have focused mainly on predicting the time to the \textit{first} default event, having ignored subsequent default events upon curing from default. While doing so is certainly expedient and simplistic, ignoring such recurrent default events also amounts to a loss of information, which may introduce excess bias into the resulting PD-estimates. In exploring this premise further, we first define a \textit{performing spell} as a multi-period episode or time span during which a bank monitors the repayment of a performing (or non-defaulted) loan at every month-end. Each performing spell has an entry time $\tau_e$ and only ends at the resolution time $\tau_r>\tau_e$, which usually coincides with the default event. 
The possibility of curing from default implies that such a loan will become subject to default risk once more; all of which implies a `multi-spell' (or recurrent event) setup for tracking loans over their lifetimes. These ideas on recurrent performing spells are illustrated in \autoref{fig:PerfSpells} for a few hypothetical loans, and across various (competing) possibilities into which a loan may resolve. Our study shall therefore have to contend with this multi-spell aspect in producing dynamic PD-estimates.

\begin{figure}[ht!]
    \centering
    \includegraphics[width=0.66\linewidth, height=0.25\textheight]{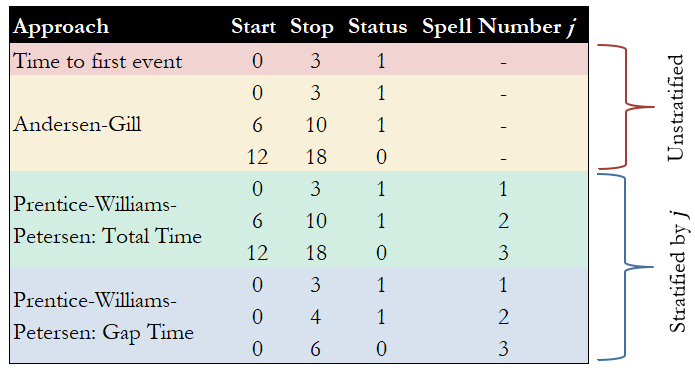}
    \caption{Illustrating the different spell-level data structures respective to each type of recurrent survival model. These data structures are shown for a hypothetical loan that defaulted twice before becoming right-censored. Inspired by \citet{ozga2018Additional}.}
    \label{fig:DataPrepRecurrent}
\end{figure}

In handling recurrent events, \citet{amorim2015modelling} and \citet{ozga2018systematic} explained that the common Cox regression model may be extended in at least two ways, each of which requires a different data layout under differing assumptions. 
Firstly, the \textit{Andersen-Gill} (AG) approach is based on increments in the number of recurrent events, as tracked by the spell number, and partitions the event history into a series of spells, whilst measuring time in terms of calendar time. The AG-approach assumes a common baseline hazard across all spells and it estimates global regression coefficients irrespective of spell number. 
Secondly, the \textit{Prentice-Williams-Peterson} (PWP) approach analyses the ordering of spells by stratifying the data based on spell number. All subjects are at risk within the first spell, though only those with a previous event in the first stratum will be at risk of the second event, and so on for successive spells. The PWP-approach can therefore incorporate both global and spell-specific effects for each covariate. Since the PWP-approach specifies a spell-specific baseline hazard function, it can also account for changes in the baseline risk between two successive spells. Moreover, the PWP-approach is usually specified in the \textit{gap time} (GT), or spell duration format, such that the time index resets to 0 at the start of each successive spell. According to \citet{kelly2000survival}, the other PWP-format is the \textit{counting process} (CP) style, or sometimes called the "total time" format, wherein the time scale reflects the time since study entry, without altering the length of time at risk within each spell.
These approaches for handling recurrent events will inform the suite of modelling techniques in our own study, and we therefore illustrate in \autoref{fig:DataPrepRecurrent} the high-level data layouts of each technique for a hypothetical loan. Regarding the PWP-technique, we shall restrict our focus to the GT-variant in this study, particularly since \citet{kelly2000survival} found negligible differences in the modelling results between the GT-- and CP-variants of the PWP-technique.

Arguably, the most common examples of multi-spell (or recurrent event) survival analysis originate 
from the biostatistical literature. From \citet{wei1997overview} and \citet[\S 8]{therneau2000modeling}, example studies hereof include recurrent infections in AIDS-patients, multiple infarcts in a coronary study, and bladder tumour recurrence after treatment.
Having used both a simulation study and an acute respiratory illness (ARI) study, \citet{kelly2000survival} compared various recurrent survival models, including the AG-- and PWP-subtypes. They found the PWP-GT variant to be superior in its ability to cater for spell-specific covariate effects and associated baseline hazards. This ability is contextually appropriate since the risk of re-contracting infections can realistically differ from spell to spell, purely given the development of immunity.
From \citet{amorim2015modelling}, the choice of technique largely depends on the maximum number of spells/events per subject, and whether the main treatment effect varies between successive spells; in which case the PWP-approach is again the more appropriate technique.
\citet{ozga2018Additional} performed a systematic investigation into the special requirements of these various techniques, having used simulation-driven studies with composite endpoints (two competing events) that may re-occur. They demonstrated that the AG-- and PWP-subtypes will resemble each other in results, but only if the baseline hazard remains relatively constant between successive spells. Should this assumption no longer hold, then the difference in results can become stark; all of which have bearing in our own study.

Literature on modelling recurrent events in credit risk is relatively limited; and even more so in developing countries, as noted by \citet{breed2021development}.
\citet{chen2012applying} investigated the determinants of upgrades/downgrades in corporate credit ratings using Standard \& Poor (S\&P) data with recurrent event Cox-models.
However, they have used the \textit{Wei-Lin-Weissfeld} (WLW) subtype in analysing these upgrade/downgrade-related endpoints, which assumes that a subject is simultaneously at risk of every event/spell. Put differently, the WLW-subtype ignores the ordering of events, and a subject can for example experience spell four without first experiencing spells one to three. 
We deem this assumption inappropriate for our context and agree with the general advice of \citet{kelly2000survival} against using the WLW-subtype when studying recurrent but ordered phenomena.
Given Zimbabwean retail loans, \citet{chamboko2016} and \citet{chamboko2019} examined the time to recovery events using a few Cox-regression subtypes, including the AG, PWP, and WLW-techniques. The work of \citet{chamboko2016} is based on a minuscule sample (4,575 obligors) with an extremely high degree of delinquency (98\%), which contrasts quite starkly with our dataset. While \citet{chamboko2019} found that the AG-technique outperformed the others, their input space was limited to only a few variables. Moreover, both studies used classical \textit{receiver operating characteristic} (ROC) analysis from \citet{fawcett2006introduction}, which can measure the discriminatory power of binary classifiers. However, this model diagnostic cannot contend with the censored nature of survival data, which affects the study results. We intend to improve upon these two studies by using \textit{time-dependent ROC} (or tROC) analysis together with a richer input space in modelling the time to default.


We address these gaps in literature by contributing a data-driven and largely pedagogical comparative study amongst three survival modelling techniques in predicting lifetime default risk. While these techniques are themselves well-established, we believe our work to be the first to benchmark them explicitly within the context of credit risk modelling. Doing so would be of value to practitioner and regulator alike when developing lifetime PD-models.
This study is accompanied by a novel suite of diagnostics, including a diagnostic tool for measuring sampling representativeness, as formulated in \autoref{sec:resolutionRate}. 
The term-structure of default risk is more rigorously defined in \autoref{sec:recurrentCoxModels} using survival analysis, whereafter we discuss and present three techniques for modelling this term-structure indirectly: 1) time to first default (TFD); 2) the AG-technique; and 3) the PWP-technique.
In appropriately assessing the discriminatory power of a Cox-model, we briefly review in \autoref{sec:ROC_time} the fundamentals of tROC-analysis in catering for the censored nature of survival data. We further adapt tROC-analysis in \autoref{sec:tROC} to deal with the dependency amongst clustered observations within each spell; a reusable contribution that we shall call the "clustered tROC-extension".
These techniques (TFD, AG, PWP) are then calibrated using South African mortgage data, as described in \autoref{sec:calibration}. Our input space contains a richer and more granular collection of time-fixed, time-varying, macroeconomic, and idiosyncratic factors; all of which engenders greater model performance.
The modelling results are themselves provided in \autoref{sec:results}, which includes a comparison of the goodness-of-fit and discriminatory power of each Cox-model. We also formulate and demonstrate a simple reusable method by which the term-structure of default risk may be estimated from such models, as inspired by survival analysis. Overall, many of these sections (particularly the reusable diagnostics) can form explicit modelling steps within a broader pedagogical tutorial for guiding the practitioner.
As such, the source code of our study is provided in \citet{botha2025recurrencySourcecode}, as implemented in the R-programming language. Finally, we conclude the study in \autoref{sec:conclusion}.

%% file: 2-Method.tex
\section{Different types of recurrent event survival models}
\label{sec:method}

Within the context of survival modelling, we formulate and discuss in \autoref{sec:resolutionRate} a diagnostic tool by which the sampling representativeness may be evaluated with respect to the raw dataset. Thereafter, three survival modelling techniques are presented in \autoref{sec:recurrentCoxModels} for modelling recurrent default events over the lifetime of loans. These techniques include \textit{time to first default} (TFD), \textit{Andersen-Gill} (AG), and \textit{Prentice-Williams-Peterson} (PWP) gap/spell-time approaches.

\input{2.1-ResolutionRate}
\input{2.2-RecurrentCoxModels}

%% file: 2.1-ResolutionRate.tex
\subsection{Testing sampling representativeness using the resolution rate \texorpdfstring{$r_\psi$}{Lg} of type \texorpdfstring{$\psi$}{Lg}}
\label{sec:resolutionRate}

In conducting survival modelling, we shall use a simple random clustered resampling scheme by which observations are randomly split into a training set $\mathcal{D}_T$ and a validation set $\mathcal{D}_V$. Loan histories are extracted in full and randomly allocated to either $\mathcal{D}_T$ or $\mathcal{D}_V$, thereby clustering around loan ID. In principle, the sets $\left\{ \mathcal{D}_T, \mathcal{D}_V \right\}$ that result from such a resampling scheme should not exhibit undue sampling bias with regard to the modelled phenomenon, especially when measured over time. However, it is unclear how exactly to measure such sampling bias within the context of survival analysis, which is why we formulate the \textit{resolution rate} as a diagnostic tool.
Consider a portfolio of $N_p$ loans, wherein any loan $i=1,\dots,N_p$ may have $j=1,\dots,n_i\geq 1$ number of performing spells. The portion of the overall loan history that is observed during each performing spell is uniquely denoted by the subject-spell construct $(i,j)$. Some spells may lack a known (or fully-observed) resolution outcome, likely due to the ongoing repayment of the loan. Accordingly, let $c_{ij}\in \{0,1\}$ indicate such right-censoring in that $c_{ij}=1$ for a right-censored spell $(i,j)$ and $c_{ij}=0$ otherwise. We refer the reader to \citet[\S 1]{kleinbaum2012survival} and \citet{schober2018survival} regarding different censoring types.

The various resolution types into which a spell $(i,j)$ may resolve can be coalesced into a single nominal variable $\mathcal{R}_{ij}$, which is encoded as 
\begin{align} \label{eq:spellResolution_Types}
    \mathcal{R}_{ij} = 
    \begin{cases}
        1: \text{Default} \quad & \text{if} \ c_{ij}=0 \ \text{and default-criteria applies} \\
        2: \text{Settled} \quad & \text{if} \ c_{ij}=0 \ \text{and settlement-criteria applies} \\
        3: \text{Write-off/Other} \quad & \text{if} \ c_{ij}=0 \ \text{and write-off (or other) criteria applies} \\        
        4: \text{Censored} \quad & \text{if} \ c_{ij} = 1
    \end{cases} \, .
\end{align}
Consider $\mathcal{R}_{ij}, i=1,\dots,N_p,j=1,\dots,n_i$ as realisations from an overall nominal random variable $\mathcal{R}$, and consider aggregating these realisations to the portfolio-level. In explaining how, let $Y_\psi\in\{0,1\}$ denote a Bernoulli random variable for a specific event type $\psi \in \mathcal{R}$. Given calendar/reporting time $t'=t_1',\dots,t_k', \dots, t_n'$, e.g., Jan-2008 to Dec-2022, assume that a series of such Bernoulli variables exist for each $\psi$, written as $Y_{\psi}(t'_1), \dots, Y_{\psi}(t'_n)$. In aggregating the realisations $\mathcal{R}_{ij}$ to the portfolio-level, let $r_\psi(t'_k,\mathcal{D})$ be the \textit{resolution rate of type} $\psi$ at which the modelled phenomenon resolves at $t'_k$ into a specified type $\psi$ within a given dataset $\mathcal{D}$. More formally, this resolution rate estimates at $t'_k$ the probability $\mathbb{P}\left(Y_\psi(t'_k) = 1 \right)$ within $\mathcal{D}$, where $r_\psi(t'_k,\mathcal{D})$ is intuitively calculated as the proportion of 1-observations of type $\psi$ in $\mathcal{D}$ at a particular time $t'$.

To aggregate these spell-level realisations $\mathcal{R}_{ij}$ towards estimating the resolution rate $r_\psi$, assume the longitudinal dataset $\mathcal{D}=\{ i,j, t_{ij}, \mathcal{R}_{ij} \}$ exists. This dataset contains categorical outcomes $\mathcal{R}_{ij}\in\mathcal{R}$ that are observed for loans $i=1,\dots,N_p$ during their respective spells $j=1,\dots,n_i$ over each spell period $t_{ij}=\tau_e,\dots,\tau_s$. Furthermore, we can partition this $\mathcal{D}$ into a series of non-overlapping monthly subsets $\mathcal{D}_s(t')$ over $t'$, where each $\mathcal{D}_s(t'_k)\in\mathcal{D}$ contains all $n_{t'_k}>0$ spells that are at risk of experiencing any event type in $\mathcal{R}$ at $t'_k$. Over each subset/cohort $\mathcal{D}_s(t')$ of size $n_{t'}$, we formally define the resolution rate $r_\psi(t',\mathcal{D})$ of type $\psi$ at each reporting time $t'=t_1',\dots,t_n'$ as 
\begin{equation} \label{eq:ResolRate}
        r_\psi(t',\mathcal{D}) = \frac{1}{n_{t'}} \sum_{(i,j) \, \in \, \mathcal{D}(t')} \mathbb{I}(\mathcal{R}_{ij}= \psi) \quad \forall \  \mathcal{D}(t') \in \mathcal{D} \ \text{and for} \ \psi \in \mathcal{R} \, ,
\end{equation}
where $\mathbb{I}(\cdot)$ is an indicator function.

The notion of reporting time $t'$ can itself be differentiated based on when spells (or cohorts thereof) commonly start or stop. In allocating spells to each monthly period $t'$, consider two contrasting definitions of tracking spell times: by spell entry time $t_e$ (or cohort-start), or by spell stop time $t_s$ (or cohort-end). Either time definition still tracks the same reporting time $t'$ in value, i.e., $t_e,t_s \in \left\{t'_1,\dots,t'_n\right\}$. The difference lies in the way in which spells are aggregated. According to the cohort-start definition, each subset $\mathcal{D}(t_e)$ contains all spells $(i,j)$ that commonly start at a particular reporting time $t'$-value, i.e., $t_e : t'=\tau_e(i,j)$. Similarly, the cohort-end definition (or spell resolution date) states that each subset $\mathcal{D}(t_s)$ includes all spells $(i,j)$ that commonly stop at a given $t'$-value, i.e., $t_s: t'=\tau_s(i,j)$. In aggregating spells, both definitions can serve two very different diagnostic purposes. The cohort-start definition can help confirm the structure of $\mathcal{D}$ in that $r_\psi$ for right-censoring ($\psi=4$) should slowly approach 100\% over time and equal 100\% at $t'_n$. More importantly, the cohort-end definition is less affected by right-censoring and can much more viably track the effect of systemic events on the portfolio (such as the 2008 global crisis), at least relative to the cohort-start definition. We shall therefore restrict our study to the cohort-end definition in the interest of expediency.

\begin{figure}[ht!]
    \centering
    \includegraphics[width=0.8\linewidth, height=0.47\textheight]{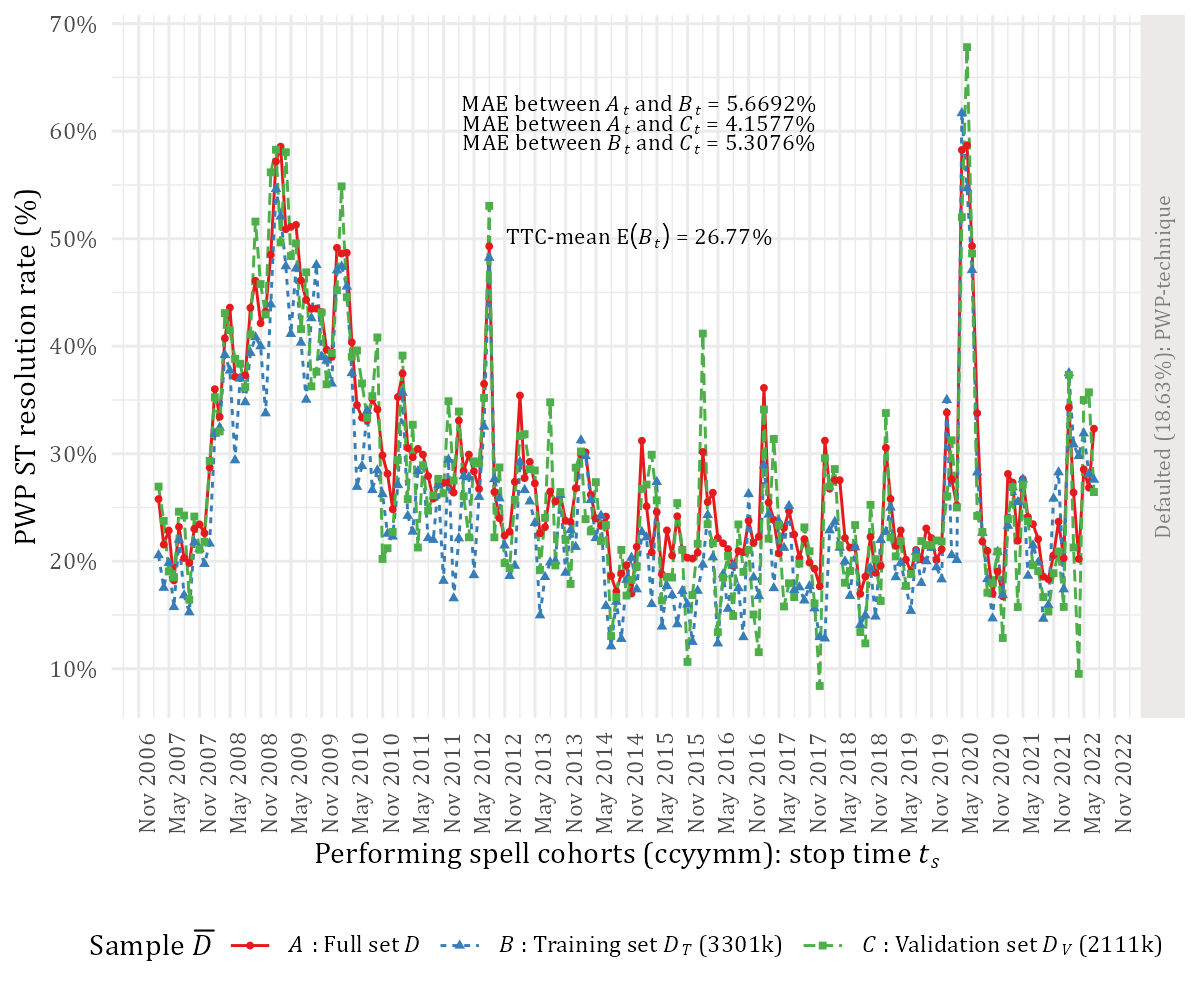}
    \caption{Comparing the resolution rates of type $\psi=1$ (Default) over time across the various datasets. The MAE-based AD-measure from \autoref{eq:ResolRate_MAE} summarises the discrepancies over time for each dataset-pair.}
    \label{fig:ResolRate_PWP}
\end{figure}

Equipped with \autoref{eq:ResolRate}, the datasets $\mathcal{D}_T$ and $\mathcal{D}_V$ can now be screened for any time-dependent sampling bias. More specifically, the resolution rates $r_\psi(t',\mathcal{D}_T)$ and $r_\psi(t',\mathcal{D}_V)$ can be duly calculated, compared, and screened for large discrepancies over reporting time $t'$. E.g., if $\mathcal{D}_T$ has an average resolution rate of 20\% for defaults but $\mathcal{D}_V$ has 5\%, then the resampling scheme may very well be deficient.
Therefore, the absolute difference, denoted as $|r_\psi(t',\mathcal{D}_T)$ - $r_\psi(t',\mathcal{D}_V)|$, should be as close to zero as possible, which would minimise sampling bias and affirm both datasets to be representative of each other. This principle suggests using the \textit{mean absolute error} (MAE) as the basis for a broader error measure in screening any two datasets against undue sampling bias. Similar to the measure used by \citet{botha2025multistate}, we define $\bar{r}_\psi\left(\mathcal{D}_1, \mathcal{D}_2\right)$ to be the \textit{average discrepancy} (AD) over calendar time $t'=t_1',\dots,t_n'$ between any two non-overlapping sets $\mathcal{D}_1$ and $\mathcal{D}_2$, expressed as the 
\begin{equation} \label{eq:ResolRate_MAE}
     \text{AD: } \quad \bar{r}_\psi\left(\mathcal{D}_1, \mathcal{D}_2 \right) = \frac{1}{n} \sum_{t'}{\big\vert r_\psi(t',\mathcal{D}_1) - r_\psi(t',\mathcal{D}_2) \big\vert} \quad \forall \ t' \ \text{and for} \ \psi \in \mathcal{R}\, .
\end{equation}
This AD-measure can be computed for all combinations of the datasets $\left\{ \mathcal{D}, \mathcal{D}_T, \mathcal{D}_V \right\}$, thereby resulting in the collection $\left\{ \Bar{r}_\psi(\mathcal{D},\mathcal{D}_T), \Bar{r}_\psi(\mathcal{D},\mathcal{D}_V), \Bar{r}_\psi(\mathcal{D}_T,\mathcal{D}_V) \right\}$ that can be compared to one another in testing sampling representativeness. We provide an example in \autoref{fig:ResolRate_PWP} for the default resolution type within the prepared datasets for the PWP-technique. Evidently, the resolution rates clearly track the 2008 financial crisis, as well as the Covid-2019 crisis. More importantly, the AD-measure confirms a visual analysis in that all resolution rates are reasonably close to one another; itself affirming low sampling bias. 
Similar results hold for those datasets of the other modelling techniques; see the codebase maintained by \citet{botha2025recurrencySourcecode}.

%% file: 2.2-RecurrentCoxModels.tex
\subsection{Term-structure of default risk: Three Cox regression models for recurrent events}
\label{sec:recurrentCoxModels}

In presenting a more mathematical definition of a `term-structure', consider $T>0$ as a discrete random variable that represents the latent lifetime of each loan $i$ with input variables $\boldsymbol{x}_i$. Then, let $h(t,\boldsymbol{x}_i)$ denote the instantaneous hazard of experiencing the default event during the discrete-valued time interval $(t-1,t]$ for loan period $t=t_{1},\dots,\mathcal{T}$, where $t_{1}$ and $\mathcal{T}$ denote respectively the time of initial recognition and ending time (or the observed lifetime). This $h(t,\boldsymbol{x}_i)$-value approximates the probability $\mathbb{P}\left(t-1 < T \leq t \, | \, T>t-1, \boldsymbol{x}_i \right)$ and represents a small `sliver' of the lifetime default probability, as discussed by \citet[pp.~17-20]{jenkins2005survival}, \citet[pp.~15-16]{crowder2012credit}, \citet{xu2016} and \citet{skoglund2017}. 
Let $S(t,\boldsymbol{x}_i)$ represent the estimated cumulative probability of each loan $i$ surviving at least up to $t$ given $\boldsymbol{x}_i$, i.e., $S(t,\boldsymbol{x}_i)$ estimates the survival probability $\mathbb{P}\left(T>t \, | \, \boldsymbol{x}_i \right)$. The function $f(t,\boldsymbol{x}_i)=S(t-1,\boldsymbol{x}_i)  h(t,\boldsymbol{x}_i)$ then represents the probability of a default event exactly at $t$, which resolves into the probability mass function $f(t,\boldsymbol{x}_i)$ in approximating $\mathbb{P}\left(T=t \, | \, \boldsymbol{x}_i\right)$ when time is discrete.
Finally, the collection $\big\{f\left(t,\boldsymbol{x}_i\right) \big\}^\mathcal{T}_{t=t_1}$ constitutes the \textit{term-structure} of default risk over loan life $t$. Note that the estimation of these quantities will be discussed later in \autoref{sec:results}.

We define the following two types of time scale definitions, respective to the two types of recurrent survival modelling techniques under consideration.
For the PWP-technique, each spell $(i,j)$ is observable from entry time $\tau_e(i,j)\geq 0$ and is recorded either up to the spell resolution time $\tau_r(i,j)$ for $c_{ij}=0$, or up to censoring time $\tau_c(i,j)<\tau_r(i,j)$ for right-censored cases $c_{ij}=1$. The overall spell stop time $\tau_s(i,j)$ is therefore simply the minimum between $\tau_r(i,j)$ and $\tau_c(i.j)$, i.e., $\tau_s(i,j) = \min{\left(\tau_r(i,j), \tau_c(i,j) \right)}$. Time is itself measured discretely during spell $(i,j)$ by the spell period $t_{ij}=\tau_e(i,j),\dots,\tau_s(i,j)$. Upon entering a new spell $j+1$, the clock is reset to $\tau_e(i,j+1)=0$ and ticks anew until reaching $\tau_s(i,j+1)$.
For the AG-technique, let $\tau_e'(i,j)\geq 0$ denote the loan period (or age) $t\in \left\{t_{i1},\dots,\mathcal{T}_i\right\}$ at which the $j^{\mathrm{th}}$ spell of loan $i$ is entered, and similarly let $\tau_s'(i,j)\geq 0$ represent the loan age at which the spell ends. Conversely, the clock keeps ticking along the lifetime of the loan under the AG-technique, with no resets.
Regardless, the overall spell age, or the total time spent thus far therein, remains the same across both techniques and we denote it as the observable failure time $T_{ij}\in\mathbb{Z}^+$, i.e., $T_{ij}=\tau_s(i,j)-\tau_e(t,j)=\tau_s'(i,j)-\tau_e'(t,j)$. Moreover, the two kinds of entry and stop times will only equal each other for the first spell, whereafter they diverge in both value and meaning.
E.g., consider a loan with two performing spells, which are recorded under the AG-technique as $t_{i1}\in(0,4]$ and $t_{i2}\in(10,13]$; whereas the spell periods are recorded as $t_{i1}\in(0,4]$ and $t_{i2}\in(0,3]$ under the PWP-technique. The corresponding data structures of all three techniques (TFD, AG, PWP) are illustrated in \autoref{app:DataStructures} for a few hypothetical loans.

For the TFD-technique, we shall use a common Cox proportional hazards model from \citet{cox1972regression}, or simply a Cox-regression, as discussed in \citet[\S 3.1]{therneau2000modeling}, \citet[\S 4.2]{crowder2012credit}
, and \citet{schober2018survival}. This model is trained only from subject-spells $(i,1)$, having excluded subsequent default events and their underlying performance spells. 
As such, we model the instantaneous hazard $h(t,\boldsymbol{x}_i)$ during time $(t-1,t]$ over $t=0,\dots,\tau_s$ for subject-spell $(i,1)$ as a function of input variables $\boldsymbol{x}_{i1}$ that are observed only for $j=1$, expressed as
\begin{equation} \label{eq:cox_TFD}
    \text{TFD: } \quad h(t,\boldsymbol{x}_{i1}) = h_0(t) \exp{\left( \boldsymbol{\beta}^\mathrm{T} \boldsymbol{x}_{i1} \right)} \, .
\end{equation}
In \autoref{eq:cox_TFD}, $h_0$ represents the baseline hazard function over $t$, $\boldsymbol{\beta}$ is a vector of estimable regression coefficients, and $\boldsymbol{x}_i=\left\{ x_{i1}, \dots, x_{ip} \right\}$ is a $p$-dimensional vector of time-fixed variables for subject-spell $(i,1)$. In fitting the TFD-model in the R-programming language, consider again the example dataset from \autoref{fig:DataPrepRecurrent}. Using the \texttt{survival}-package, the TFD-model is fit by first creating a \texttt{Surv}-object as an outcome variable of sorts, which specifies certain timing-related fields within a given survival dataset, as well as maps the event indicator field. Thereafter, the \texttt{coxph()}-function relates a set of covariates to this \texttt{Surv}-object, resulting in a \texttt{coxph}-object. We illustrate its coding as follows.
\begin{lstlisting}
    modTFD <- coxph( Surv(Start, Stop, event=Status==1) ~ Covariates, data=datTFD, id=Spell_Key)
\end{lstlisting}

An important concept when estimating $\boldsymbol{\beta}$ is the idea of a \textit{risk set}, which differs across our recurrent event models. For a series of $m$ unique ordered failure times $t_{(1)}<,\dots,<t_{(k)}<,\dots,<t_{(m)}$, let $R\left(t_{(k)}\right)$ be the risk set at failure time $t_{(k)}$, as explained by \citet[pp.~62--63,~81--82]{crowder2012credit}. Each risk set $R\left(t_{(k)}\right)$ contains all those subject-spells that are at risk of the main event just prior to $t_{(k)}$, which includes both those spells that experienced the event at $t_{(k)}$, as well as those that were lost to right-censoring at $t_{(k)}$. Risk sets are particularly useful when assembling the partial likelihood function towards estimating $\boldsymbol{\beta}$, since these risk sets determine which risk scores $\exp{\left( \boldsymbol{\beta}^\mathrm{T} \boldsymbol{x}_{i} \right)}$ are summed together across certain spells; see \citet{kelly2000survival}, \citet[\S 3.1]{therneau2000modeling}, and \citet[pp.~62--63]{crowder2012credit}. Specifically, the partial likelihood for the Cox model from \autoref{eq:cox_TFD} is expressed for $j=1$ across all unique failure times as
\begin{equation} \label{eq:partialLikelihood}
    \mathrm{PL}(\boldsymbol{\beta}) = \prod_{q=1}^m{ \left( \frac{ \exp{ \left( \boldsymbol{\beta}^\mathrm{T}\boldsymbol{x}_{ij} \right)}  }{\sum_{(i,j) \, \in \, R\left(t_{(q)}\right)} { \exp{ \left( \boldsymbol{\beta}^\mathrm{T}\boldsymbol{x}_{ij} \right)} } } \right) } \, ,
\end{equation}
where the risk set is defined as $R(t) = \left\{ (i,j) :  \tau_s(i,j) \geq t  \right\}$ for $j=1$ and $i=1,\dots,N_p$.

By definition, the AG-technique assumes a common baseline hazard $h_0$ across all spells, which implies the Cox model
\begin{equation} \label{eq:cox_AG}
    \text{AG: } \quad h(t,\boldsymbol{x}_{ij}) = h_0(t) \exp{\left( \boldsymbol{\beta}^\mathrm{T} \boldsymbol{x}_{ij} \right)} \, .
\end{equation}
The partial likelihood from \autoref{eq:partialLikelihood} remains the same, though it is now estimated across all spells $j=1,\dots,n_i$ of each loan $i$. Furthermore, the risk set is slightly different in that its constituent spells are those that are at-risk between the calendar-based stop times of spell $j-1$ and $j$, i.e., $R(t) = \left\{ (i,j) : \tau_s'(i,j-1) <  t \leq \tau_s'(i,j)   \right\}$. Given the common baseline hazard, any spell-specific effect (if it exists) can only enter the Cox model via the input variables. The AG-model is implemented in R in exactly the same fashion as the TFD-model, except for specifying a different dataset (e.g., \texttt{datAG} instead of \texttt{datTFD}) that has the required data layout; itself shown in \autoref{fig:DataPrepRecurrent} and \autoref{app:DataStructures}.

The Cox model under the PWP-technique incorporates a spell-specific baseline hazard $h_{0j}$ for each spell $j=1,\dots, J$ up to the maximum observed spell $J$. By implication, the model specification becomes
\begin{equation} \label{eq:cox_PWP}
    \text{PWP: } \quad h(t,\boldsymbol{x}_{ij}) = h_{0j}(t) \exp{\left( \boldsymbol{\beta}^\mathrm{T} \boldsymbol{x}_{ij} \right)} \, .
\end{equation}
As before, the partial likelihood from \autoref{eq:partialLikelihood} remains unchanged from that of the AG-technique, though the risks set is now defined using gap time lengths (or spell ages), i.e., $R(t) = \left\{ (i,j) : T_{ij} \geq t \right\}$. For more in-depth explanations of the differences in these risk sets, see the discussions by \citet{kelly2000survival} and \citet{ozga2018Additional}.
Lastly, the PWP-model is implemented in R by specifying the strata-argument to be the spell number $j$, illustrated as follows.
\begin{lstlisting}
    modPWP <- coxph( Surv(Start, Stop, Status==1) ~ Covariates + strata(SpellNum), data=datPWP, id=Spell_Key)
\end{lstlisting}

%% file: 3-tROC.tex
\section{Time-dependent ROC-analysis for survival models}

In \autoref{sec:ROC_time}, we briefly review the fundamentals of time-dependent ROC-analysis (or "tROC") towards measuring the discriminatory power of a Cox regression model. Thereafter, a new extension to tROC-analysis is presented in \autoref{sec:tROC}, which can contend with the dependence structure amongst observations in our survival data.

\input{3.1-ROC_time}

\input{3.2-tROC}

%% file: 3.1-ROC_time.tex
\subsection{A brief review of classical time-dependent ROC-analysis}
\label{sec:ROC_time}

In comparing the performance of different survival models, a key question is that of their ability to discriminate accurately amongst accounts at a higher/lower risk of the event. A \textit{receiver operating characteristic} (ROC) curve is a traditional and popular form of analysis to evaluate the discriminatory power of a binary classifier; see \citet{fawcett2006introduction}. This ROC-curve graphs the trade-off between the true positive rate $T^+$ and the false positive rate $F^+$. However, a classical ROC-curve cannot truly measure a survival model's discriminatory power since some of the observations are still pending due to right-censoring. Since the time frame $t\geq 0$ can vary across which the survival probability is predicted, an ROC-based test of discriminatory power will also vary in tandem with the degree of right-censoring over $t$. As a remedy, \citet{heagerty2000} and \citet{bansal2018tutorial} showed $T^+$ and $F^+$ to be functions of time. Consider a random variable $M$ that represents the marker values (or risk scores) from a Cox-model, such that greater values of $M$ denote greater risk of the event, and vice versa. Let the random variable $T$ denote the latent lifetime of subjects $s=1,\dots,n$, and let $D(t)$ be a generic counting process in that $D(t)=1$ if $T\leq t$ for a failed subject and $D(t)=0$ otherwise. In rendering predictions, we need to dichotomise $M$ using a variable threshold $p_c$, i.e., a positive event if $M>p_c$ and a negative event otherwise. Thereafter, one can plot $T^+$ against $F^+$ in following the \textit{cumulative cases / dynamic controls} (CD) approach, expressed respectively as 
\begin{align} \label{eq:TPR}
    T^+(p_c,t) &= \mathbb{P}\big(M>p_c \, | \, T \leq t \big) = \mathbb{P}\big(M>p_c \, | \, D(t)=1\big) \, , \quad \text{and} \\ \label{eq:FPR}
    F^+(p_c,t) &= 1-\mathbb{P}\big(M\leq p_c \, | \, T > t \big) = 1-\mathbb{P}\big(M \leq p_c \, | \, D(t)=0\big) = \mathbb{P}\big(M>p_c \, | \, D(t)=0 \big) \, .
\end{align}
In following \citet{heagerty2000}, one may rewrite \crefrange{eq:TPR}{eq:FPR} using the conditional survivor function $S(t \, | \, M>p_c)$ that is estimated only within the subset of those cases with markers $M>p_c$. Using Bayes' theorem, it follows that \crefrange{eq:TPR}{eq:FPR} can be rewritten respectively as
\begin{align} \label{eq:tpc_bayes}
    T^+(p_c,t) &= \frac{\big(1-S\left(t \rvert \, M>p_c\right)\big) \mathbb{P}(M>p_c)}{1-S(t)}  \\  \label{eq:fpc_bayes}
    F^+(p_c,t) &= 1-\frac{S\left(t\rvert \, M \leq p_c\right) \mathbb{P}(M \leq p_c)}{S(t)} \, .
\end{align}

While a few approaches exist for estimating $S(t)$, we shall restrict our review to the \textit{Nearest Neighbour} (NN) estimator given its favourable properties, as originally proposed by \citet{akritas1994} and explored by \citet{heagerty2000}. In particular, the NN-estimator calculates the bivariate survivor function $S(p_c,t)= \mathbb{P}(M>p_c,T>t)$ up to a given prediction time $t$, and is expressed as the
\begin{equation}
\label{eq:S_NN_estimator_usual}
    \text{Akritas-estimator:} \quad \hat{S}_{\lambda_n} (p_c,t) = \frac{1}{n} \sum_{s=1}^n \hat{S}_{\lambda_n} \big(t \, \rvert \,M = m_s \big) \mathbb{I} \big(m_s > p_c \big) \, .
\end{equation} 
\autoref{eq:S_NN_estimator_usual} represents the average (conditional) survivor function across those marker values that exceed the given $p_c$-threshold, where $\lambda_n$ is a smoothing parameter; itself discussed later. 
In defining the \textit{marker-conditional} survivor function $S\big(t \rvert M = m_s \big)$ in \autoref{eq:S_NN_estimator_usual}, first consider the unique failure time vector $\boldsymbol{t}$, along with the following quantities that correspond to each marker $m_s,s=1,\dots,n$: the subject age $T_s$, and the resolution type $\mathcal{R}_s$ that resolves to 1 if the main event occurred at a given time $t$ and 0 otherwise. 
Assuming that smoothing will be required, consider the sequence of all time-ordered markers $m_1,\dots,m_w,\dots,m_n$, where the ordering is based on the subject ages $T_{(1)}< \cdots<T_{(w)}<\cdots < T_{(n)}$. For a given marker $m_w$, the authors then defined the smoothed estimator $\hat{S}_{\lambda_n}\big(t \, \rvert \, M = m_{w} \big)$ using the NN-related kernel function $K_{\lambda_n}\in\{0,1\}$ as a weight within a Kaplan-Meier type estimator, expressed as 
\begin{equation}
\label{eq:S_NNE_weightedKM}
    \hat{S}_{\lambda_n} \big(t \, \rvert \, M = m_w \big) = \prod_{q \, \in \, \boldsymbol{t}, \ q \leq t} \left\{ 1 - \frac{\sum_{s=1}^n K_{\lambda_n}\left(m_{s}, m_w\right) \mathbb{I} \left(T_{(s)} = q, \mathcal{R}_{s}=1\right)} {\sum_{s=1}^{n} K_{\lambda_n}\left(m_s, m_w\right) \mathbb{I}  \left(T_{(s)} \geq q\right)} \right\} \, .
\end{equation}

Regarding the smoothing weights in \autoref{eq:S_NNE_weightedKM}, consider a kernel function $K_{\lambda_n}(m_{s},m_{w})$ with corresponding smoothing parameter $\lambda_n$ across any pair of ordered marker values $(m_{s},m_{w})$ for $s\ne w$.
\citet{akritas1994} specifically examined a $0/1$ nearest neighbour kernel function (hence the name "NN"), defined as 
\begin{equation} \label{eq:NN_kernel}
    K_{\lambda_n}\left(m_{s},m_w\right) = \mathbb{I}\left(- v(\lambda_n) < \hat{F}_M\left(m_{s}\right) - \hat{F}_M\left(m_w\right) < v(\lambda_n) \right)  \, ,
\end{equation}
which produces a weight $k_s\in\{0,1\}$ corresponding to each $(m_{s},m_w)$. In \autoref{eq:NN_kernel}, $\hat{F}_M(m_s)$ is the empirical marker distribution, and $2\lambda_n \in (0,1)$ is the proportion of observations included within each neighbourhood. Moreover, each neighbourhood is bounded by $[-v(\lambda_n),v(\lambda_n)]$, where $v(\cdot)$ produces a neighbourhood bound from the underlying sequence of time-ordered markers.
Put differently, $k_s=1$ indicates that marker $m_s$ is within the neighbourhood of (or sufficiently similar to) the slightly larger $m_w$, and vice versa for $k_s=0$.
While other kernel choices are certainly possible, \citet{heagerty2000} noted that any NN-kernel will result in ROC-estimates that are invariant to monotone transformations of $M$. 
Finally, and assuming that $T$ and $M$ are mutually independent, $T^+$ and $F^+$ from \crefrange{eq:tpc_bayes}{eq:fpc_bayes} are updated respectively using conditional probability as 
\begin{equation} \label{eq:tpr_NNE}
    T^+(p_c,t) = \frac{\mathbb{P}(M>p_c) - \mathbb{P}(T>t \, | \, M>p_c)\mathbb{P}(M>p_c)}{1-S(t)} =\frac{\big(1-\hat F_{M} (p_c)\big) - \hat S _{\lambda_n} (p_c,t)}{1-\hat S_{\lambda_n} (t)} \, \text{;} \quad \text{and}
\end{equation} 
\begin{equation} \label{eq:fpr_NNE}
     F^+(p_c,t) = \frac{\mathbb{P}(T>t)\big( \mathbb{P}(M\leq p_c) + \mathbb{P}(M>p_c)  \big) - \mathbb{P}(T>t\, | \,M\leq p_c)\mathbb{P}(M\leq p_c)}{S(t)}
     =\frac{\hat S_{\lambda_n} (p_c,t)}{\hat S_{\lambda_n} (t)},
\end{equation} where $\hat S_{\lambda_n} (t) = \hat S_{\lambda_n} (p_c,t)$ for the boundary cut-off value of $p_c=-\infty$.

%% file: 3.2-tROC.tex
\subsection{Dealing with clustered observations: The clustered tROC-extension}
\label{sec:tROC}

It is yet unclear how the NN-estimator from \citet{heagerty2000} will fare when dealing with observations in survival data that are clustered around certain spells (or subjects). In particular, consider the markers $m_{ijt}\in M$ over spell periods $t_{ij}=1,\dots,T_{ij}$, where the $t$ in $m_{ijt}$ represents $t_{ij}$ as a simplification of notation. These markers are clustered around a particular spell $(i,j)$ of loan $i=1,\dots,N_p$, itself spanning $j=1,\dots,n_i$ spells. This data structure clearly embeds an explicit dependency in that certain markers are explicitly related to a specific spell of a loan; rows are therefore not necessarily independent from one another. 
In contrast, the NN-estimator clearly assumes that $m_s,s=1,\dots,n$ are independent from one another. It may very well be that this NN-estimator will produce biased results when failing to account for the aforementioned dependency structure.

As a possible remedy, one may treat the clustered markers $m_{ijt}$ as subpopulations of the spells $(i,j)$, whereafter certain quantities within the NN-estimator can be reformulated accordingly using arithmetic means. Consider that the Akritas-estimator from \autoref{eq:S_NN_estimator_usual} is by definition the average value of the estimated bivariate survivor function $\hat{S}(p_c,t)$-values over $n$ markers.
Given the pre-existing use of the arithmetic mean within the Akritas-estimator, the markers $m_{ijt}$ may be similarly summarised across the $T_{ij}$ spell periods for each $(i,j)$, before their incorporation into the Akritas-estimator. 
More formally, we redefine the Akritas-estimator as the \textit{mean-adjusted Akritas} (MAA) estimator for a given threshold $p_c$ and time horizon $t$, denoted as $\hat{S}_{\lambda_n}^{b} (p_c,t)$ and expressed as
\begin{equation}
\label{eq:S_NN_estimator_meanAdj}
    \text{mean-adjusted Akritas-estimator:} \quad 
    \hat{S}_{\lambda_n}^{b} (p_c,t) = \frac{1}{n} \sum_{(i,j)} \left\{
    \frac{1}{\eta_{ij}}\sum_{v=1}^{T_{ij}} 
    \hat{S}_{\lambda_n} \big(t \, \rvert \,M = m_v \big) \mathbbm{I} \big(m_{v} > p_c \big) \right\} \, .
\end{equation}
In \autoref{eq:S_NN_estimator_meanAdj}, $n$ is the number of subject-spells $(i,j)$ over which we shall take the average, and $\eta_{ij}$ denotes the risk set size of those qualifying marker values $m_v=m_{ijt}$, i.e., those markers over $t_{ij}=1,\dots,T_{ij}$ where $m_{ijt} > p_c$.

Similarly, one may adjust the estimator $\hat{F}_M(m)$ of the cumulative marker distribution $F_M(m)$ towards aligning with the structure of the MAA-estimator in \autoref{eq:S_NN_estimator_meanAdj}, which now operates at the spell-level instead of marker-level. This implies taking the spell-level average $m_v$ of those markers $m_{ijt}$ for $t_{ij}=v=1,\dots, T_{ij}$ of each $(i,j)$ that are at or below a given cut-off, whereafter the average is taken again across these estimates. We define this quantity as the
\begin{equation} \label{eq:marker_CDF_meanAdj}
    \text{mean-adjusted marker distribution:} \quad
    \hat F_M (m)= \frac{1}{n} \sum_{(i,j)} \left\{ \frac{1}{\eta_{ij}} \sum_{v=1}^{T_{ij}} \mathbbm{1}(m_v \leq m) \right\} \, .
\end{equation}
We shall label the application of these mean-adjusted quantities from \crefrange{eq:S_NN_estimator_meanAdj}{eq:marker_CDF_meanAdj} as the "clustered tROC-extension" of time-dependent ROC-analysis. An application hereof (called the \texttt{tROC.multi()}-function) is developed in R, which can be found in the codebase maintained by \citet{botha2025recurrencySourcecode} in script 0b(iii).
Future work can surely investigate the statistical properties of this clustered tROC-extension; our focus remains on conducting a pedagogical benchmark study.

%% file: 4-DataCalibration.tex
\section{Calibrating the recurrent event Cox regression models to mortgage data}
\label{sec:calibration}

Our survival models are trained using a data-rich portfolio of residential mortgages from a large South African bank. This longitudinal dataset contains $47,942,462$ monthly observations relating to the repayment performance of each loan $i=1,\dots,N_p; N_p=653,317$ across its particular lifetime. The portfolio and its constituents are observed from January 2007 up to December 2022, during which time new mortgages were continuously originated every month. Left-truncated loans, i.e., those loans whose performance predates the starting date of the study, are retained with all of their available histories. This dataset not only contains a rich input space for predictive modelling, but also contain fundamental credit fields, e.g., net cash flows (receipts), expected instalments, arrears balances, month-end balances, interest rates, loan principals, and the amount and timing of write-offs and early settlement.
This rather large dataset $\mathcal{D}$ is subsampled into a smaller but still representative sample $\mathcal{D}_S\in \mathcal{D}$. Apart from attaining greater computational expediency, doing so offsets the adverse effect of large sample sizes on $p$-values when testing the statistical significance of regression coefficients; a point discussed by \citet{lin2013LargeSamples}. As such, we employ stratified clustered random sampling by extracting from $\mathcal{D}$ the full credit histories of 90,000 loans, based on balancing greater sample sizes against computational effort. Loan keys are randomly selected within each stratum, where strata are based on the loan status, i.e., a completed, active, settled, or written-off loan. Of these 90,000 loans, 70\% are randomly relegated within each stratum into the training set $\mathcal{D}_T\in\mathcal{D}_S$, whilst sorting the remainder into the validation set $\mathcal{D}_V\in\mathcal{D}_S$.
While the $\mathcal{D}_T$-set remains unchanged for most techniques, we remove recurrent spells $j\geq 2$ from $\mathcal{D}_T$ for the TFD-technique only.
Regarding data preparation tasks, we: 1) rectify zero-valued starting balances and loan principals; 2) treat unflagged account closures; 3) fix illogical event amounts at loan termination; and 4) employ the TruEnd-procedure from \citet{botha2025truEnd} towards identifying and discarding trailing zero-valued (or very small) balances. See the codebase by \citet{botha2025recurrencySourcecode} for details.

\begin{figure}[ht!]
    \centering
    \includegraphics[width=0.7\linewidth, height=0.4\textheight]{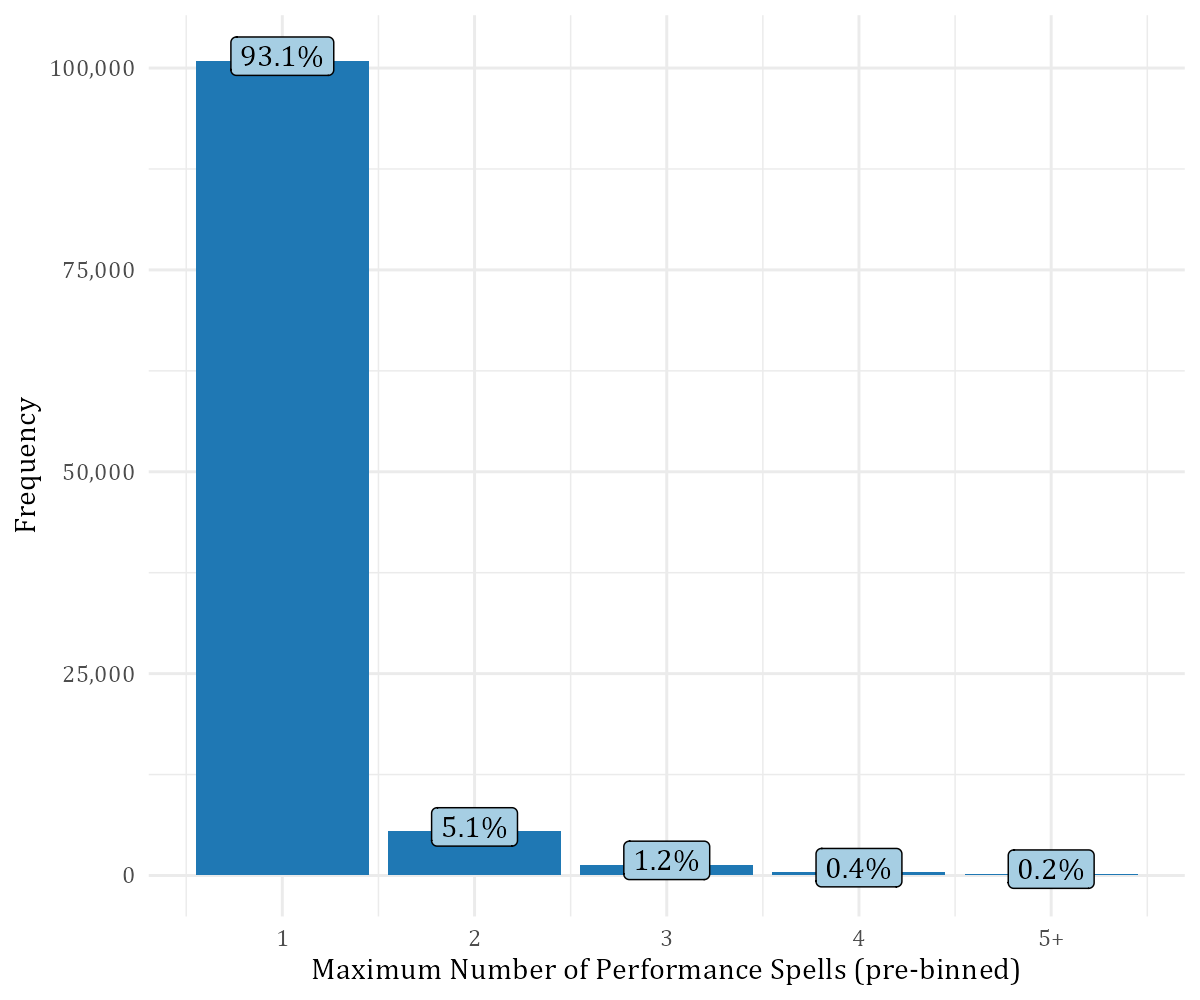}
    \caption{Distribution of the maximum number of performance spells experienced per loan, drawn from the full set $\mathcal{D}$.}
    \label{fig:MaxPerfSpellNum}
\end{figure}

\begin{figure}[ht!]
    \centering
    \includegraphics[width=0.7\linewidth, height=0.4\textheight]{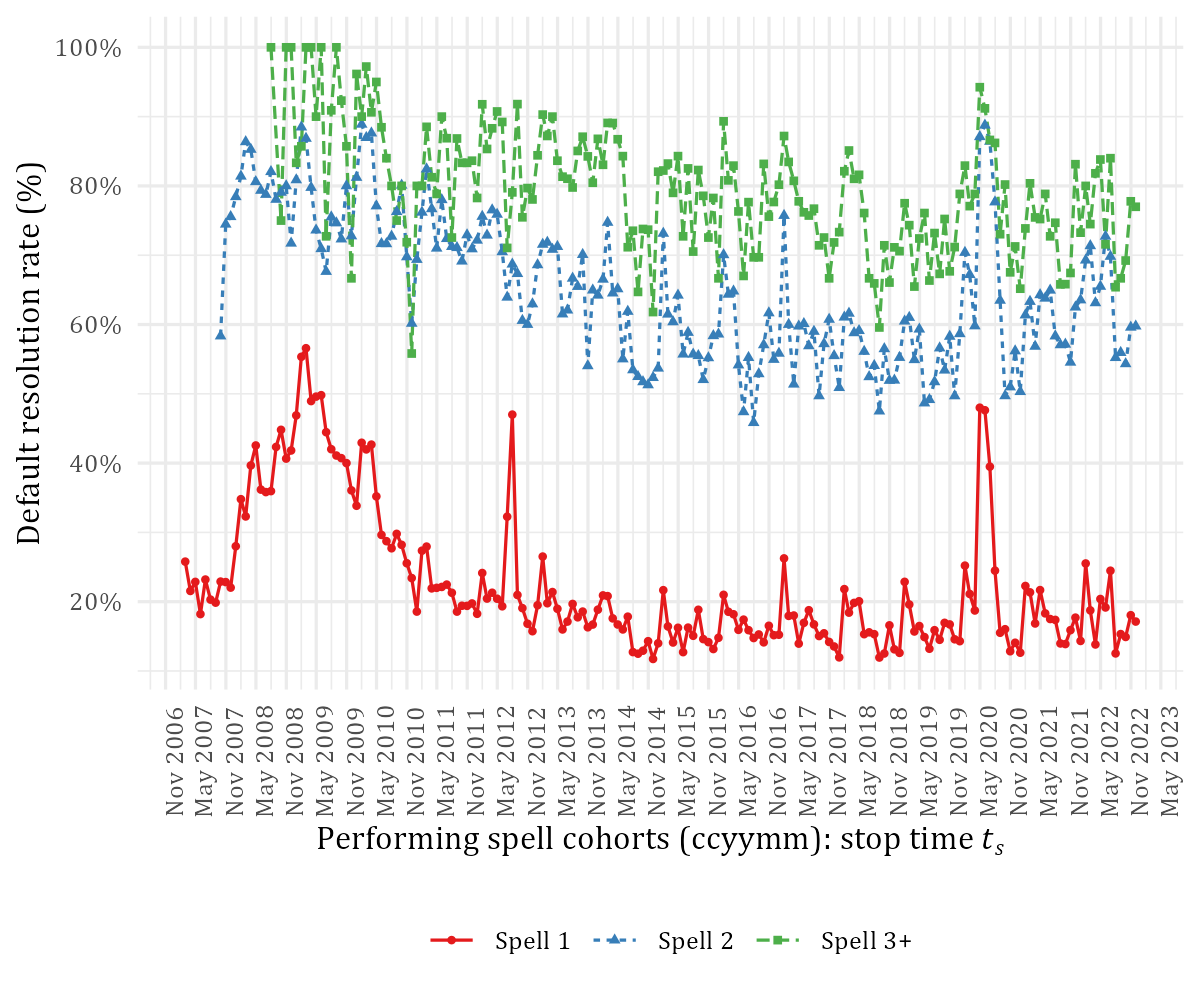}
    \caption{Resolution rate $r_\psi(t')$ of type $\psi=1$ (Default) over reporting time $t'$, calculated per numbered spell and using the cohort-end $t_s$ time scale. Spell numbers beyond three are grouped together simply for graphical fidelity.}
    \label{fig:DefResolRate_SpellNumber}
\end{figure}

In determining the degree of recurrent events, we graph in \autoref{fig:MaxPerfSpellNum} the number of observed maximum spells per loan. Whilst the vast majority of cases ($\approx 93\%$) only had a single performance spell, we argue that the remaining multi-spell cases are sufficiently prevalent in warranting a multi-spell modelling method. Similar to \citet{chamboko2019}, a few borrowers in our dataset have experienced up to ten performing spells. However, and given the dwindling sample sizes at these later spells, we shall bin together all numbered spells beyond four. Since one might also bin data towards isolating different behaviours, we graph in \autoref{fig:DefResolRate_SpellNumber} the different default resolution rates over time, as grouped by spell number. Evidently, the default experience changes markedly per spell increment, which further corroborates the premise of our study.

In fitting any model, an oft-overlooked yet crucial step is the selection of input variables, as lamented by \citet{heinze2018variable} and \citet{Kakarla2021}. This step is similar to a football coach selecting players—evaluating skills, minimizing redundancy, and refining the final team. Likewise, we follow a thematic variable selection process using repeated Cox-regressions across themed subsets of input variables. This rather interactive process is guided by the following aspects that we shall use as `tools' in screening each input within each final model. 
Firstly, and as a working principle, we strive towards attaining model parsimony by using the smallest number of inputs relative to the sample size, thereby aiding human interpretability. This goal is balanced against achieving the maximum goodness-of-fit value, as discussed by \citet{Akaike1998} and measured using the \textit{Akaike Information Criterion} (AIC).
Secondly, we use domain expertise in screening variables and structuring them into various themes, e.g., delinquency, loan-level characteristics, portfolio-level inputs, and macroeconomic covariates. Each theme has an overarching question, e.g., \textit{"which lagged version of the policy/repurchase rate is `best' in predicting the outcome?"}, that is ultimately answered using the aforementioned aspects/tools. 
Thirdly, rank-based correlation studies (Spearman) are conducted across each themed subset towards identifying clusters of correlated variables. A cluster can prompt dividing the constituent variables into further subthemes for testing.
Fourthly, we test the statistical significance of each input using the Wald-statistic against a significance level of $\alpha=0.05$.
Fifthly, we measure the in-sample goodness-of-fit of each final model using median-adjusted Cox-Snell (CS) residuals, which ought to follow a unit exponential distribution. As devised by \citet{ansin2015evaluation}, the degree to which the CS-residuals deviate from the unit exponential is evaluated by using the test statistic $D$ of a two-sample Kolmogorov-Smirnov test. Greater values of $1-D$ indicate smaller departures from the assumption, and hence a better fit.
Lastly, we assess the discriminatory power of each input variable when used within a single-factor Cox-model, as will be described shortly.
All of these insights are then collated across themes, thereby forming a combined input space that is itself curated further using domain expertise. 
For greater detail on our thematic selection process, see the comments within the codebase by \citet{botha2025recurrencySourcecode}. This process culminates in a unique input space per recurrent event Cox-model, as summarised in \autoref{app:InputSpace}.

\begin{figure}[ht!]
    \centering
    \includegraphics[width=0.7\linewidth, height=0.4\textheight]{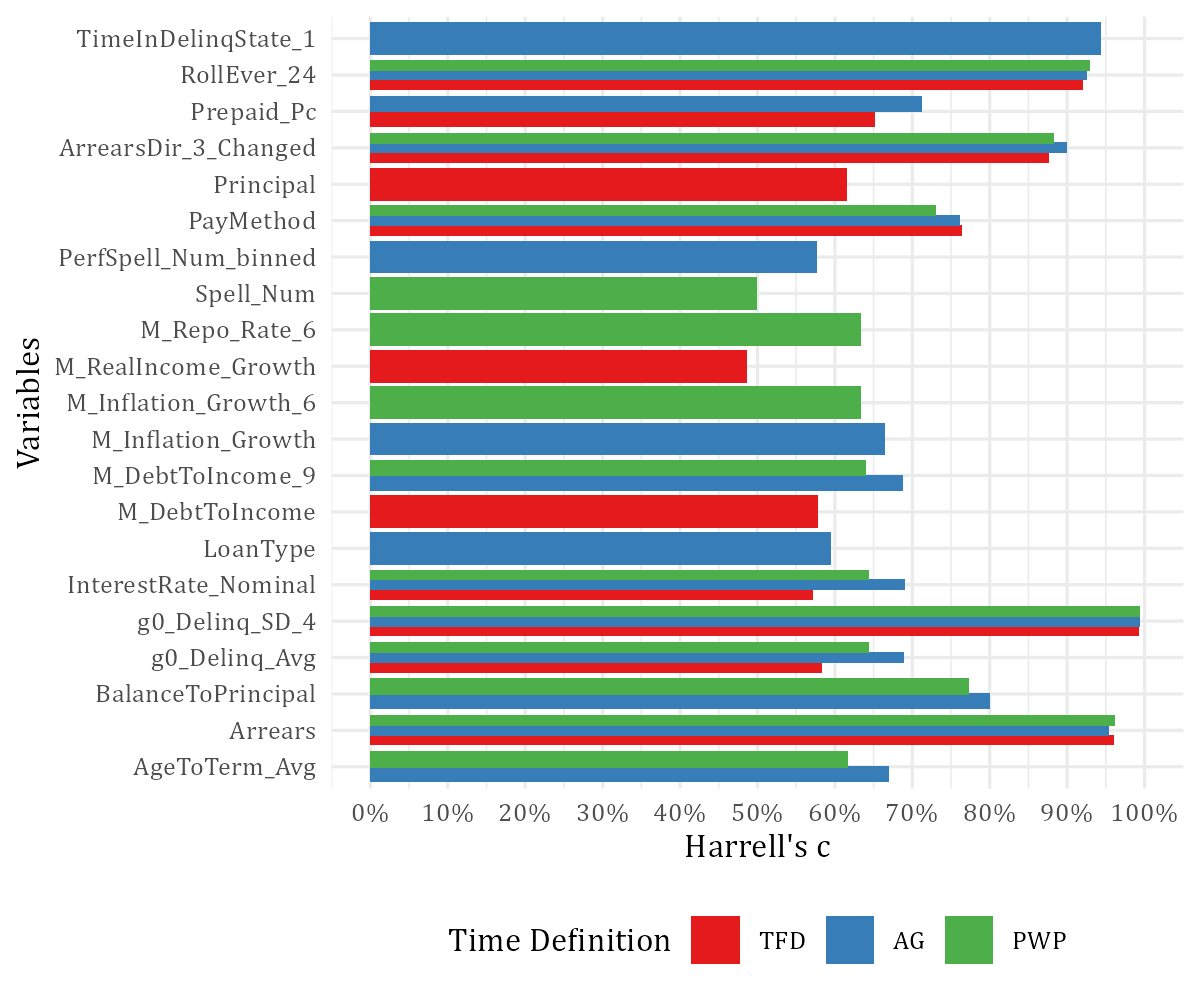}
    \caption{Comparing the Harrell $c$-statistics of single-factor models across all three types of recurrent event models.}
    \label{fig:HarrellC}
\end{figure}

As part of our thematic selection process, we measure the extent to which any input variable contributes to the discriminatory power of a Cox regression model. Accordingly, we construct various single-factor (or single-variable) Cox-regressions within each subtheme, and summarise their performance using Harrell's $c$-statistic (or concordance). As explained by \citet{gonen2005concordance} and \citet{royston2013CoxValidation}, the $c$-statistic is the proportion of spell pairs in which the predictions and outcomes are concordant within survival data; greater $c$-values indicate better discriminatory power. We ascribe the greatest importance to the variable whose single-factor model has the greatest $c$-statistic.
These $c$-statistics are reported in \autoref{fig:HarrellC} per variable, having used the finalised input space per recurrent event technique. Although the results can vary significantly across techniques, there is a clear trend in that delinquency-themed variables have the greatest $c$-values (and hence importance), e.g., the variables \texttt{g0\_Delinq\_SD\_4}, \texttt{ArrearsDir\_3\_Changed}, and \texttt{Arrears}. This finding suggests quite intuitively that the practitioner should at least include delinquency-themed variables when building Cox-models in analysing the time to default.

%% file: 5-Results.tex
\section{Comparing different recurrent event Cox-models across various diagnostics}
\label{sec:results}

\begin{figure}[ht!]
    \centering
    \begin{subfigure}{0.49\textwidth}
        \caption{Goodness-of-fit: TFD-technique} \label{fig:KS_statistic_a}
        \centering\includegraphics[width=1\linewidth,height=0.27\textheight]{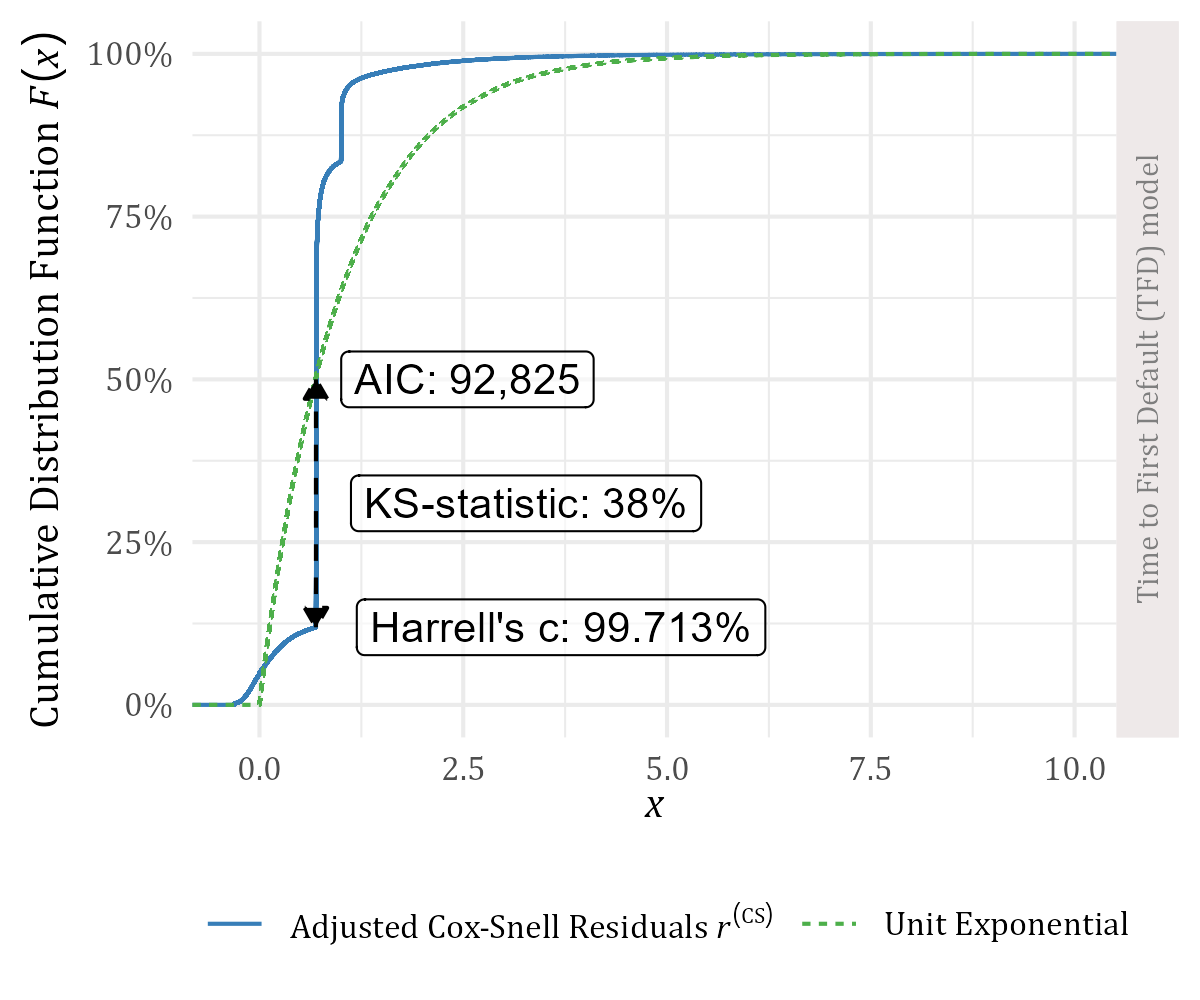}
    \end{subfigure}
    \begin{subfigure}{0.49\textwidth}
        \caption{Goodness-of-fit: AG-technique} \label{fig:KS_statistic_b}
        \centering\includegraphics[width=1\linewidth,height=0.27\textheight]{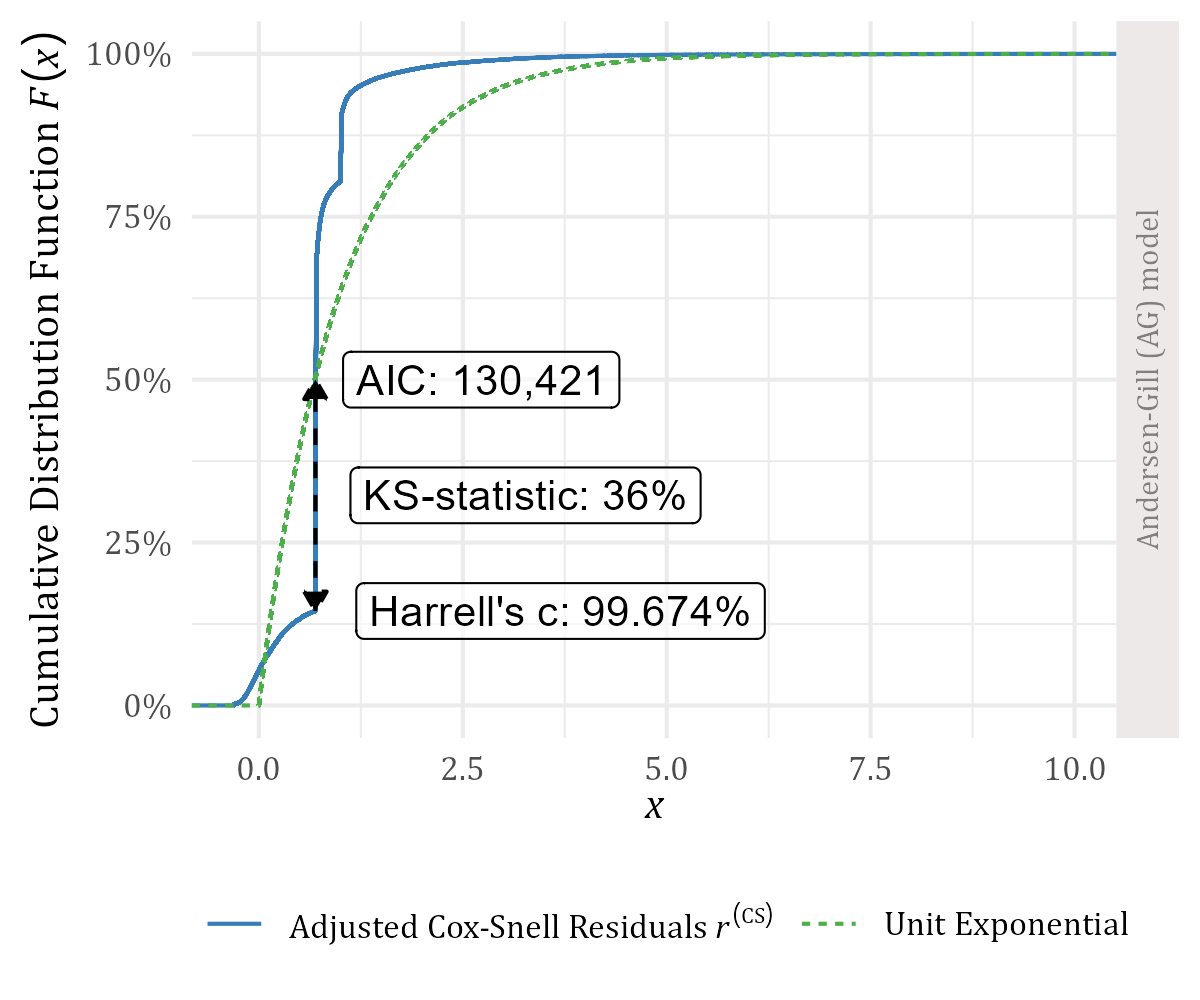}
    \end{subfigure}
    \begin{subfigure}{0.50\textwidth}
        \caption{Goodness-of-fit: PWP-technique} \label{fig:KS_statistic_c}
        \centering\includegraphics[width=1\linewidth,height=0.27\textheight]{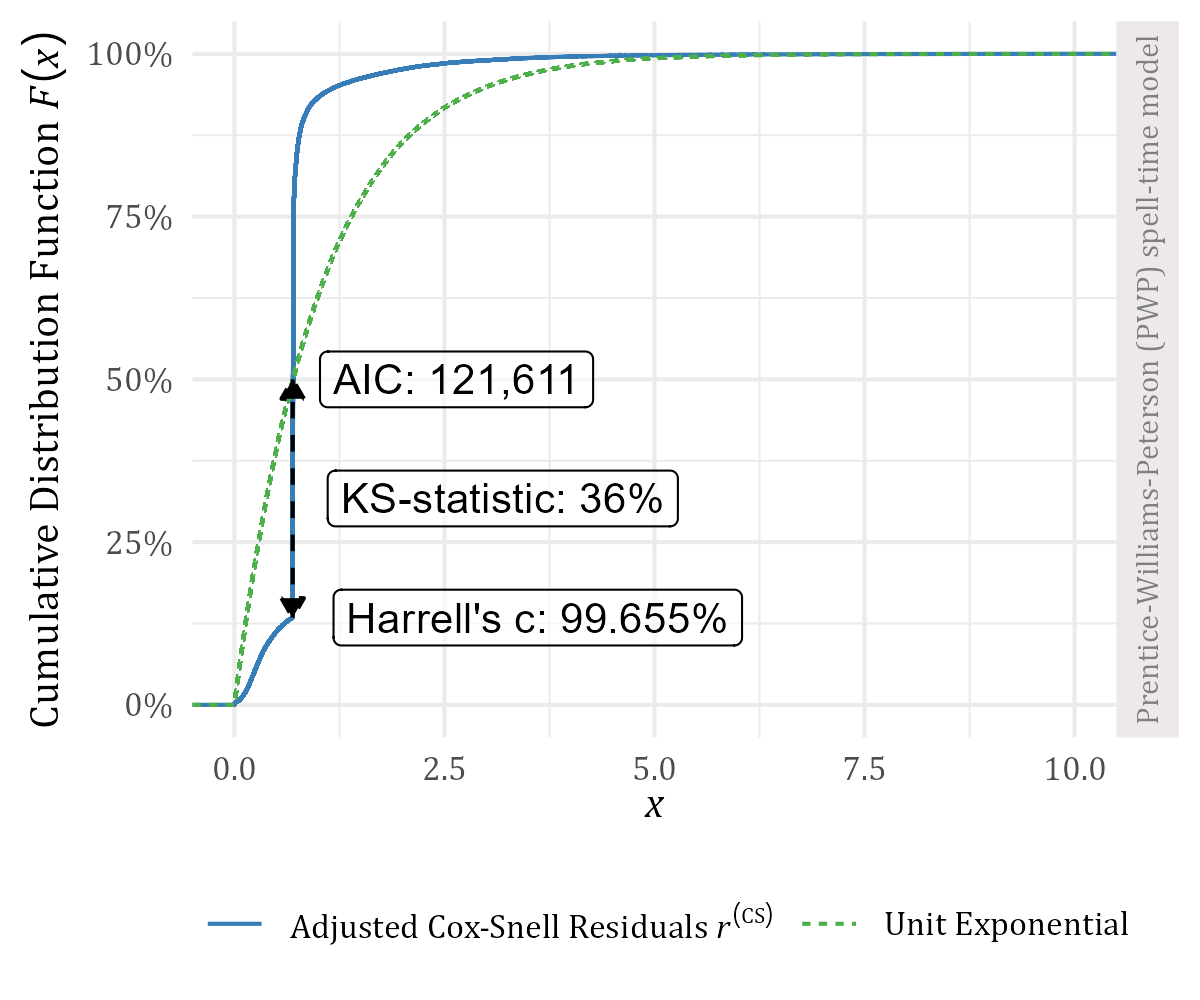}
    \end{subfigure}    
    \caption{Testing the goodness-of-fit of each Cox regression model by comparing the distribution of the median-adjusted Cox-Snell (CS) residuals against a unit exponential distribution, respective to each modelling technique in panels (\textbf{a})--(\textbf{c}). Overlaid statistics include the AIC, Harrell's $c$, and the KS test statistic $D$.}\label{fig:KS_statistic}
\end{figure}

We shall evaluate our Cox regression models across two critical aspects of any modelling exercise: goodness-of-fit (GoF) and discriminatory power. Measures that evaluate these aspects are fundamental in assessing the quality of fit to training data, as well as the degree to which the model can render accurate predictions beyond the training data. Firstly, and in measuring GoF, we present in \autoref{fig:KS_statistic} the cumulative distributions of the median-adjusted Cox-Snell (CS) residuals that arise from each recurrent event Cox-model: TFD, AG, and PWP. Evidently, there is little difference in the quality of fit amongst all three techniques, in that each residual distribution resembles (to a certain degree) the unit exponential distribution. The degree of this discrepancy between either distribution is measured using the Kolmogorov-Smirnov test statistic $D$, and we note that all techniques have similar $D$-values.
Secondly, Harrell's $c$-statistic is calculated for each model in measuring its discriminatory power: 99.713\% (TFD), 99.674\% (AG), and 99.655\% (PWP). This result suggests that the TFD-technique has an ever so slightly better ability in distinguishing between those spells that experience the default-event and those that do not. Admittedly, these $c$-statistics are all extremely high, though we ascribe the superior discriminatory power to the quality of input variables, as well as to the process by which they are selected into the various Cox-models.

\begin{figure}[ht!]
    \centering
    \begin{subfigure}{0.49\textwidth}
        \caption{tROC-analysis: TFD-technique} \label{fig:tROC_a}
        \centering\includegraphics[width=1\linewidth,height=0.27\textheight]{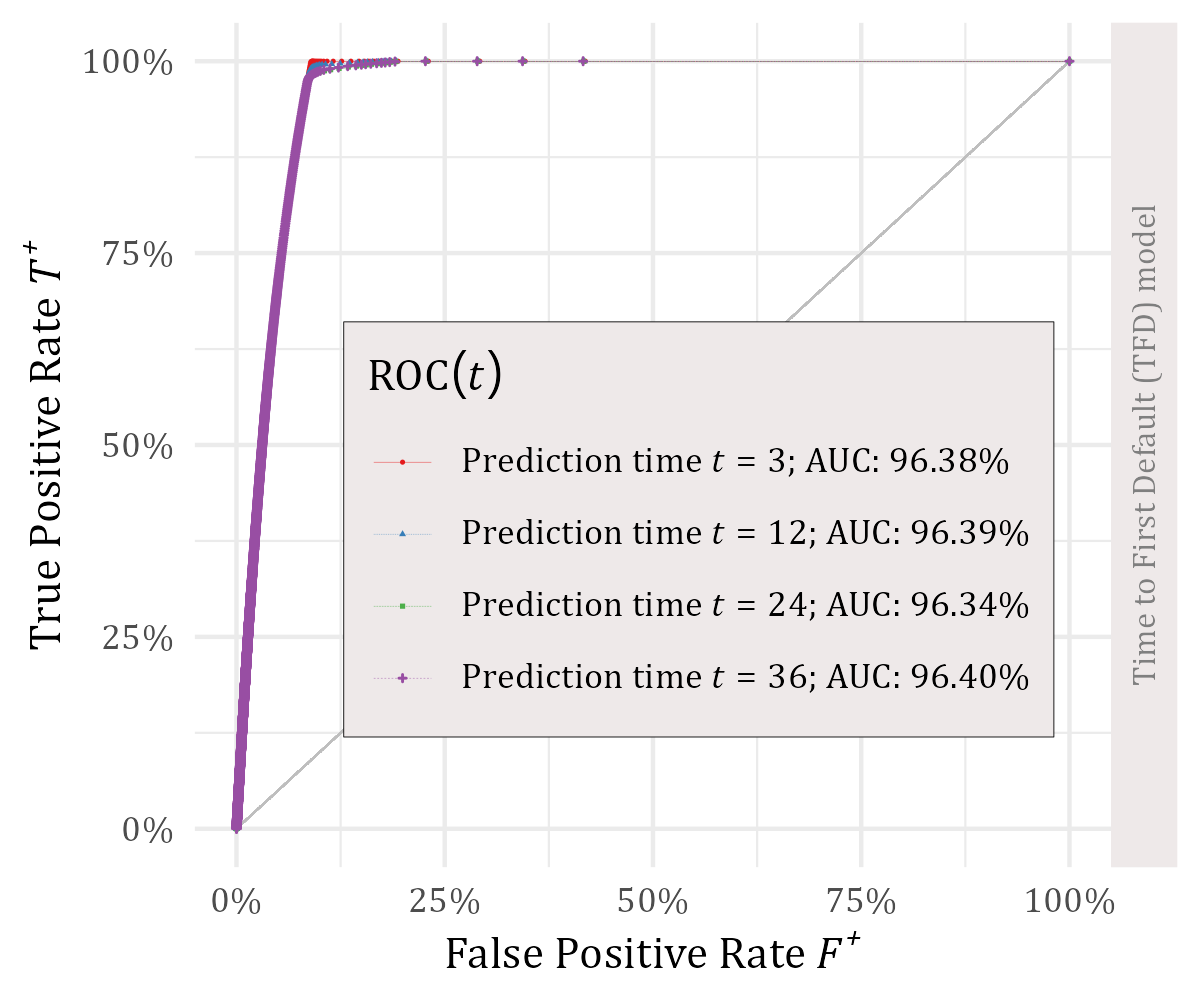}
    \end{subfigure}
    \begin{subfigure}{0.49\textwidth}
        \caption{tROC-analysis: AG-technique} \label{fig:tROC_b}
        \centering\includegraphics[width=1\linewidth,height=0.27\textheight]{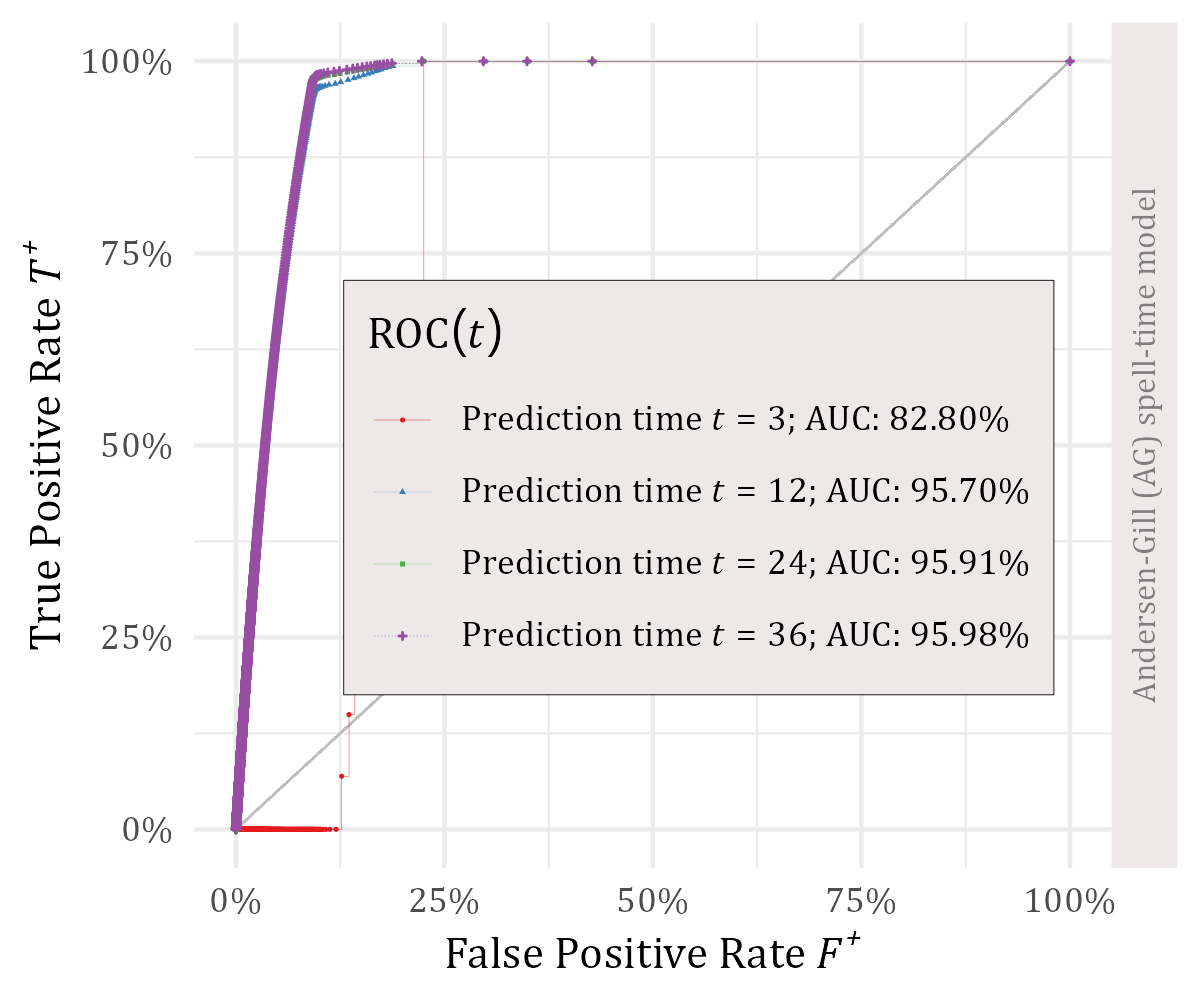}
    \end{subfigure}
    \begin{subfigure}{0.50\textwidth}
        \caption{tROC-analysis: PWP-technique} \label{fig:tROC_c}
        \centering\includegraphics[width=1\linewidth,height=0.27\textheight]{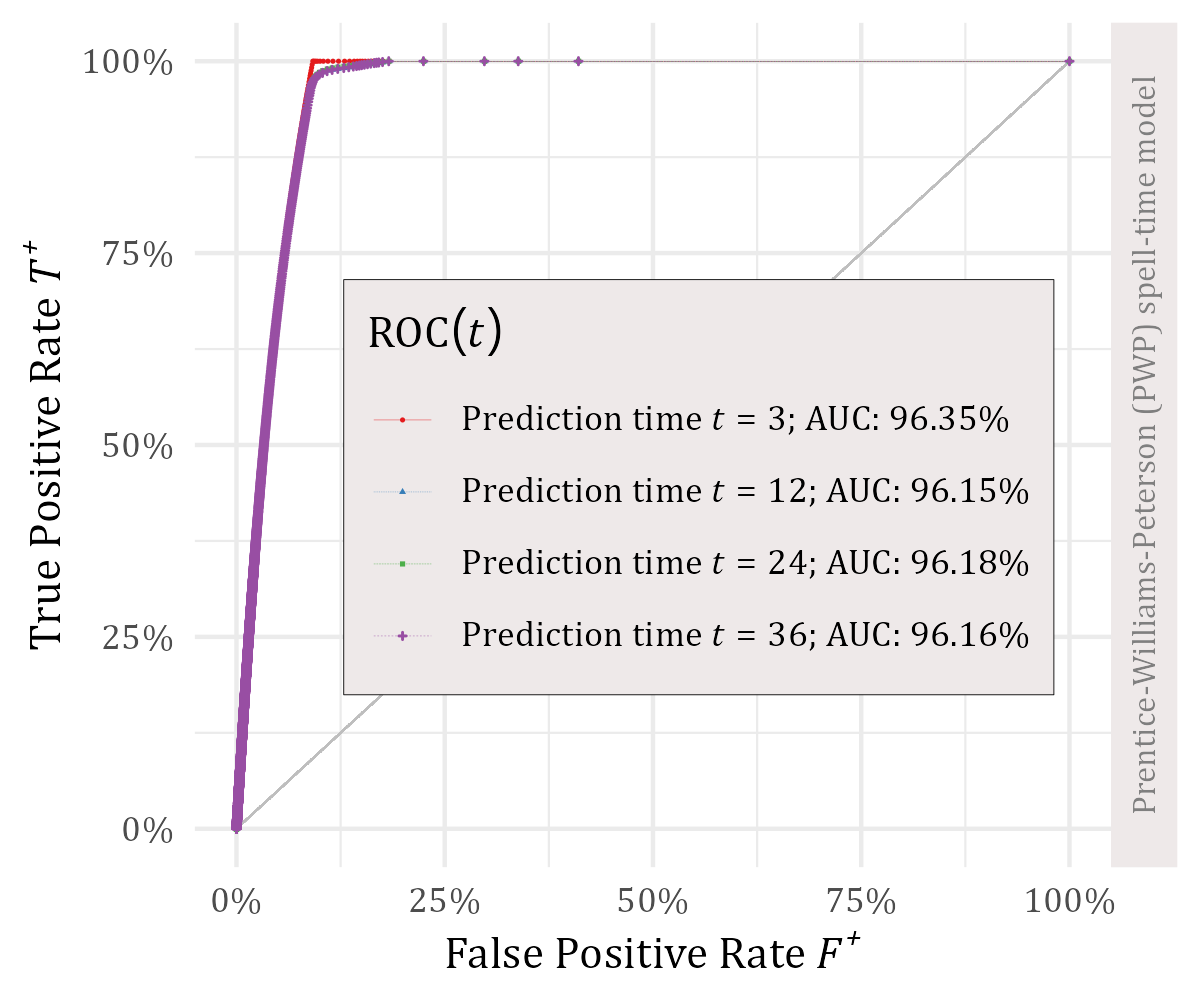}
    \end{subfigure}    
    \caption{Testing the discriminatory power of each Cox regression model by applying the clustered tROC-extension from \autoref{sec:tROC} of time-dependent ROC-analysis, respective to each technique in panels (\textbf{a})--(\textbf{c}). Time horizons include three, twelve, twenty-four, and thirty-six months.}\label{fig:tROC}
\end{figure}

The discriminatory power of these Cox regression models may also be assessed over specific time horizons by using our clustered tROC-extension from \autoref{sec:tROC}. These tROC-analyses are provided in \autoref{fig:tROC}, which are summarised into a single quantity per time horizon by using the \textit{time-dependent area under the curve} (tAUC) statistic; itself printed in \autoref{fig:tROC}. Greater values of tAUC indicate stronger discriminatory power for a given time horizon. We do not observe any discernable trend in tAUC-values across longer time horizons for any specific technique. This result is surprising at first since predictions rendered over longer outcome periods are usually less accurate than those over shorter periods. In fact, this result does not corroborate the works of \citet{kennedy2013window} and \citet{botha2025sicr}, which studied the effect of the outcome period on discriminatory power using cross-sectional models (i.e., logistic regression). However, the near-constant tAUC-values might attest to the inherent ability of survival models to render quality predictions across any time horizon; a trait that is not shared by cross-sectional models.
Furthermore, the tAUC-values are remarkably close to one another across technique, despite the intrinsic benefits of either the AG-- or PWP-techniques. That said, the PWP-technique does seem to outperform the AG-technique ever so slightly, whereas the TFD-technique has a slight edge in performance over the PWP-technique.
Overall, these high tAUC-values suggest that no real benefit exists when opting for any specific recurrent event Cox-model over another, at least from the perspective of discriminatory power.

Aside from GoF and discriminatory power, we evaluate the ability of these Cox-models to generate a term-structure of PD-estimates over all time horizons. 
Consider the actual term-structure, denoted by $\big\{f_\mathrm{A}\left(t\right) \big\}^\mathcal{T}_{t=t_1}$ over unique default times $t$ during loan life, as introduced in \autoref{sec:recurrentCoxModels}. Assuming discrete time, the default event probability $f(t)=\mathbb{P}(T=t)$, or marginal PD at a given $t$, is derived using the Kaplan-Meier (KM) estimator $\hat{S}(t)$ of the survival probability $\mathbb{P}(T>t)$. Together with $\hat{S}(t)$, we define the elements of the actual term-structure over $t$ as 
\begin{equation} \label{eq:eventProb_actual}
    f_\mathrm{A}(t)=\hat{S}(t-1)\hat{h}(t) \, ,
\end{equation}
where $\hat{h}(t)=d_t/n_t$ is the associated hazard rate, $d_t$ is the number of defaults, and $n_t$ is the number of at-risk spells at each $t$. In the R-programming language, the KM-estimator is implemented within the \texttt{survfit()}-function when given a normal \texttt{Surv}-object, as discussed in \autoref{sec:recurrentCoxModels}. This function produces an overall survival curve $\hat{S}$ with associated $\hat{h}$-estimates over $t$. From these estimates, one can then derive $f_\mathrm{A}$ from \autoref{eq:eventProb_actual} at each $t$, thereby resulting in the actual term-structure of default risk. See script 4a(ii) in the R-codebase from \citet{botha2025recurrencySourcecode} for details. We do however provide a high-level code snippet below of the KM-estimator and the event probability, as calculated for a given dataset (\texttt{dat}); itself specific to a particular technique (TFD, AG, or PWP).

\begin{lstlisting}
    modKM <- survfit( Surv(Start, Stop, Status==1) ~ 1, data=dat, id=Spell_Key)
    datSurv <- surv_summary(modKM) %>% as.data.table() # survival table
    datSurv[, Hazard := n.event / n.risk]
    datSurv[, EventRate := shift(surv, n=1, fill=1) * Hazard]
\end{lstlisting}

The predicted survival probability $\hat{S}\left(t,\boldsymbol{x}_{ij}\right)$ may be similarly obtained from a fitted Cox-model given a particular spell $(i,j)$ and its set of time-dependent covariates $\boldsymbol{x}_{ij}$. Practically, we again use the \texttt{survfit()}-function, though this time with a fitted \texttt{coxph}-object (which represents a specific Cox-model) to obtain individual predicted survival curves for each spell over its duration. Greater detail is given in script 5c in the R-codebase from \citet{botha2025recurrencySourcecode}, whereas the fitting of a \texttt{coxph}-object was discussed in \autoref{sec:recurrentCoxModels}. As before, we provide a cursory code snippet below, which scores the survival and the associated event probability of a single spell $(i,j)$ over its discrete-time periods $t_{ij}$.
\begin{lstlisting}
    datSpell <- subset(datPWP, Spell_Key = unique(datPWP$Spell_Key)[1])
    objSurv <- survfit(modPWP, centered=F, newdata=datSpell, id=Spell_Key)
    datSurv <- data.table(Time=datSpell$Stop, surv=objSurv$surv)
    datSurv[, Survival_1 := shift(Survival, n=1, type="lag", fill=1]
    datSurv[, Hazard := (Survival_1 - Survival) / Survival_1]
    datSurv[, EventRate := Survival_1 * Hazard]
\end{lstlisting}
Obtaining such survival curves is a computationally-intensive process that spans many hours since \texttt{survfit()} is called once for each spell, despite being called within a multithreaded environment. Nonetheless, we are able to derive the default event probabilities over time, expressed for a single spell $(i,j)$ as
\begin{equation} \label{eq:eventProb_exp}
    f_\mathrm{P}\left(t, \boldsymbol{x}_{ij}\right)=\hat{S}(t-1\, | \, \boldsymbol{x}_{ij})\hat{h}(t,\boldsymbol{x}_{ij}) \, .
\end{equation}
Note that $\hat{h}(t,\boldsymbol{x}_{ij})$ in \autoref{eq:eventProb_exp} is itself approximated using $\hat{S}(t-1\, | \, \boldsymbol{x}_{ij})$, expressed as
\begin{equation}
 \hat{h}(t,\boldsymbol{x}_{ij}) = \frac{\hat{S}(t-1\, | \, \boldsymbol{x}_{ij}) - \hat{S}(t\, | \, \boldsymbol{x}_{ij})}{\hat{S}(t-1\, | \, \boldsymbol{x}_{ij})} \, .
\end{equation}

In compiling the expected term-structure of a portfolio, we posit that one may take the arithmetic average of the subject-level $f_\mathrm{P}\left(t, \boldsymbol{x}_{ij}\right)$-estimates at each spell period $t$, i.e., the average event probability
\begin{equation}
    f_\mathrm{P}(t) = \frac{1}{n_t}\sum_{(i,j)}{f_\mathrm{P}\left(t, \boldsymbol{x}_{ij}\right)} \, .
\end{equation}
The resulting term-structure $\big\{f_\mathrm{P}\left(t\right) \big\}^\mathcal{T}_{t=t_1}$ may then be compared to the actual variant $\big\{f_\mathrm{A}\left(t\right) \big\}^\mathcal{T}_{t=t_1}$ towards evaluating each model's accuracy at the portfolio-level. The average discrepancy between either term-structure can be quantified by calculating the \textit{mean absolute error} (MAE) over $t$, expressed as
\begin{equation} \label{eq:mae_termStructure}
    \\\frac{1}{\mathcal{T}-t_1} \sum_{t=t_{1}}^{\mathcal{T}}{\big\vert f_{\mathrm{A}}(t)-f_{\mathrm{P}}(t) \big\vert}
     \, .
\end{equation}

\begin{figure}[ht!]
    \centering\includegraphics[width=0.7\linewidth,height=0.4\textheight]{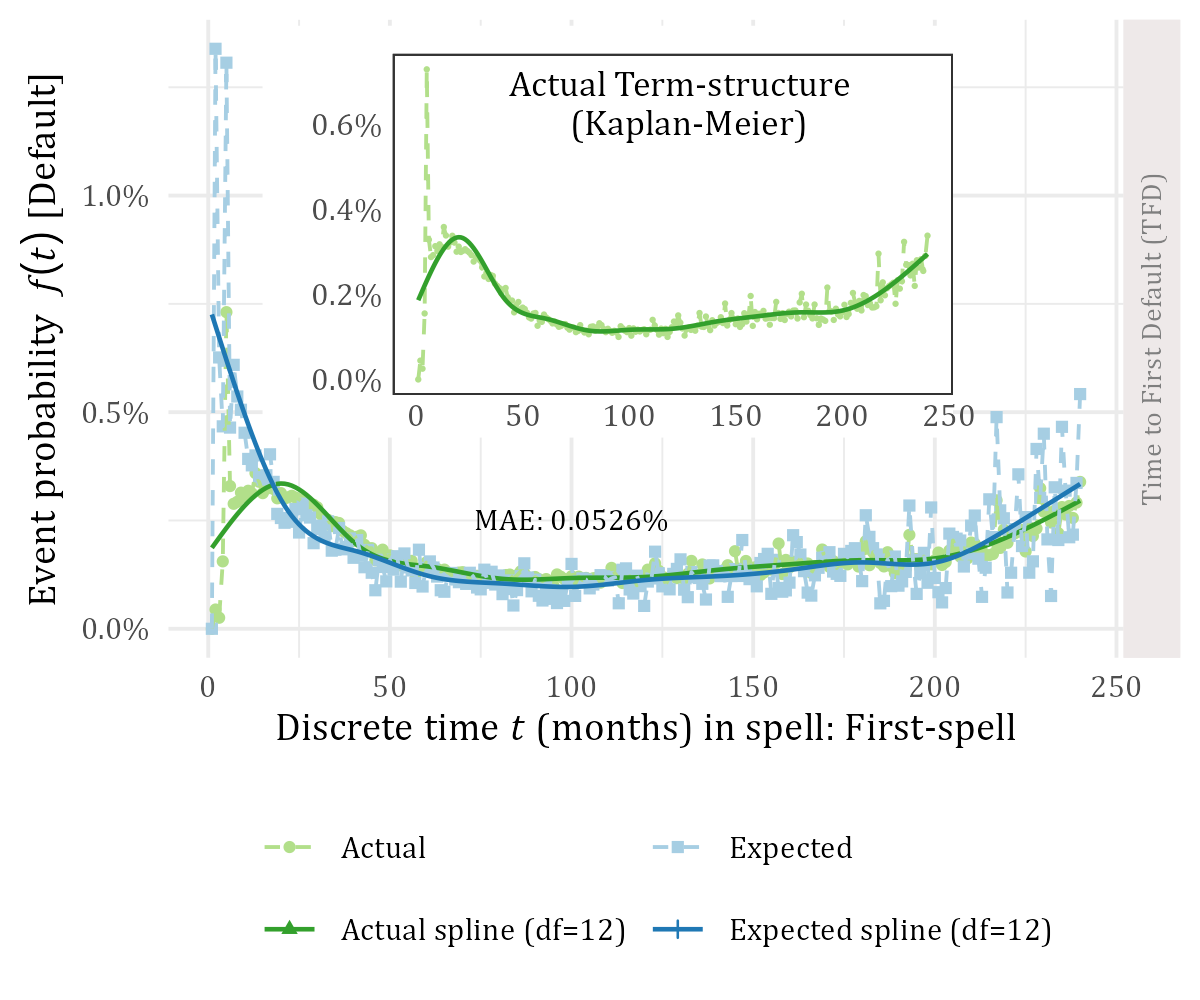}
    \caption{Comparing the actual vs expected average event probability over time in spell, i.e., the term-structure of default risk, respective to the TFD-technique. Natural cubic splines are overlaid simply to illustrate the general trend. The MAE from \autoref{eq:mae_termStructure} summarises the average discrepancy between the actual and expected cases. An inset graph shows the actual term-structure in isolation, merely due to scaling in the main graph.}\label{fig:TermStructure_TFD}
\end{figure}

\begin{figure}[ht!]
    \centering\includegraphics[width=0.7\linewidth,height=0.4\textheight]{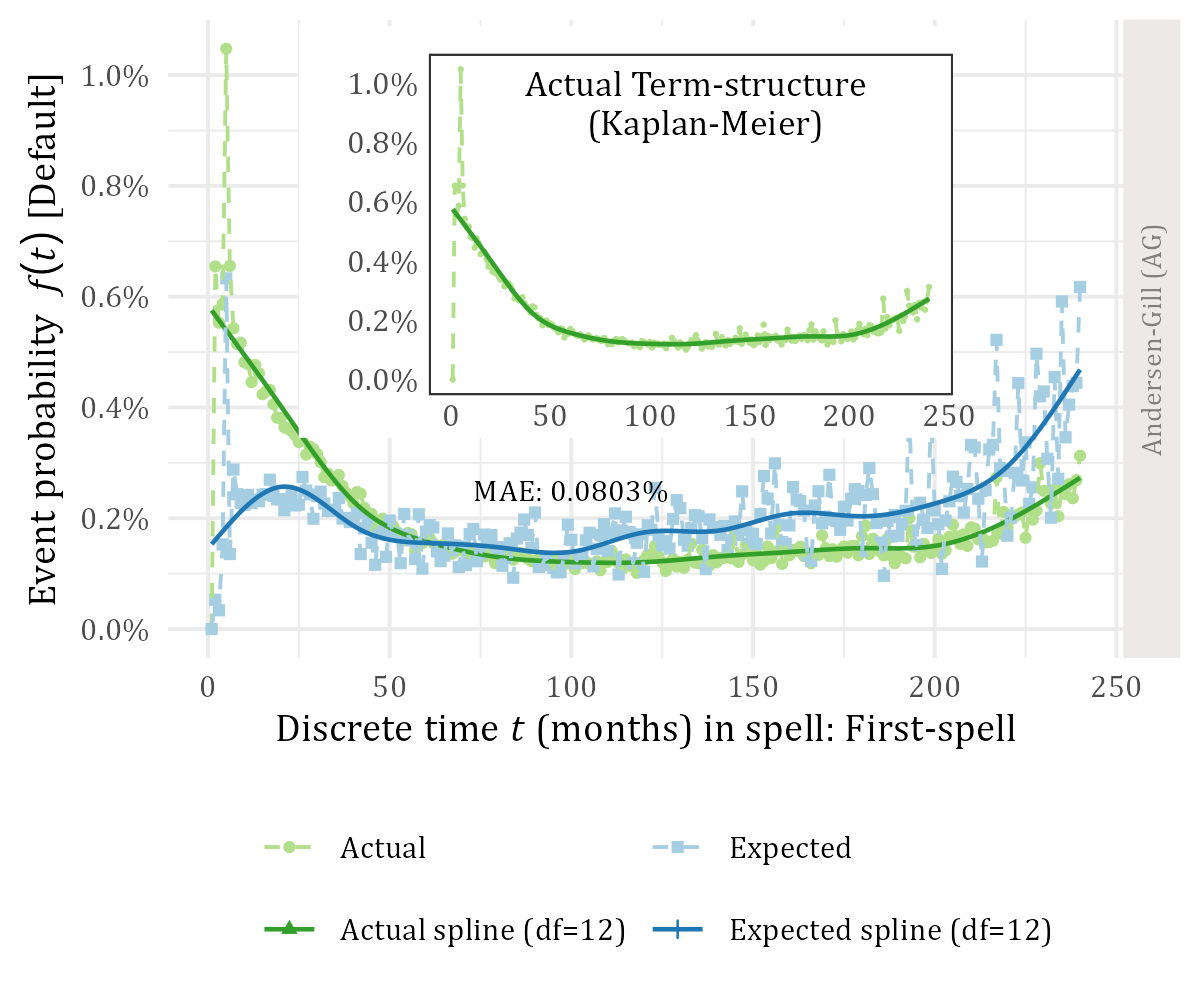}
    \caption{Comparing the actual vs expected average event probability over time in spell, i.e., the term-structure of default risk, respective to the AG-technique. Graph design follows that of \autoref{fig:TermStructure_TFD}}\label{fig:TermStructure_AG}
\end{figure}

\begin{figure}[ht!]
    \centering\includegraphics[width=0.7\linewidth,height=0.4\textheight]{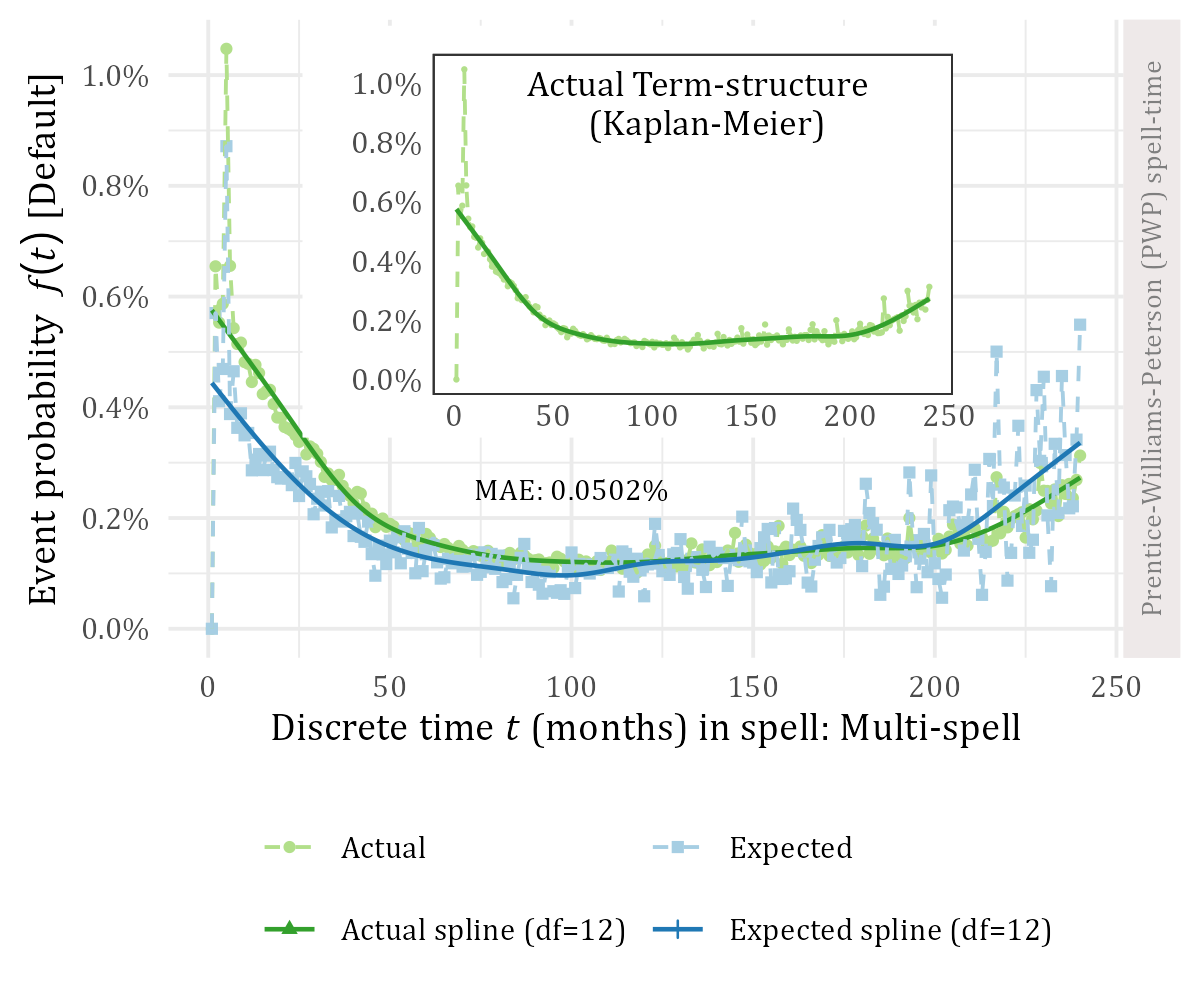}
    \caption{Comparing the actual vs expected average event probability over time in spell, i.e., the term-structure of default risk, respective to the PWP-technique. Graph design follows that of \autoref{fig:TermStructure_TFD}}\label{fig:TermStructure_PWP}
\end{figure}

We present the actual-expected term-structure graphs in \crefrange{fig:TermStructure_TFD}{fig:TermStructure_PWP} respective to each recurrent event technique. 
The spell periods $t$ are limited to a maximum of 240 months, which not only enhances graphical fidelity, but also recognises that the vast majority (99.99\%) of the dataset is exhausted at this point.
Furthermore, the term-structures for the TFD-technique include only the first performance spell by definition. Excluding subsequent spells would explain why the actual term-structure differs slightly in \autoref{fig:TermStructure_TFD} from those term-structures in \crefrange{fig:TermStructure_AG}{fig:TermStructure_PWP}, respective to the AG-- and PWP-techniques. 
Nonetheless, $f_\mathrm{A}$ still exhibits a "U-shape" over $t$ in all of the actual term-structures, which agrees with industry experience. At first, default risk is high during the earlier periods of spell lifetimes, whereafter it gradually subsides as the relationship between bank and borrower is "worn-in" regarding loan repayment. Later parts of spell life have slightly elevated levels of default risk, particularly so for the TFD-technique. This insurgence of defaults is largely attributed to early settlements and/or mortgage sales, which are predicated by strategic defaults during the sale; itself often a lengthy process. The sample size also progressively dwindles during these later parts of spell life, which explains the increased volatility in all estimates of $f_\mathrm{A}(t)$ during those later periods $t\geq 175$.

We further find that all expected term-structures approximate their actual counterparts quite reasonably across all techniques, at least so for most periods of spell lifetimes. This result indicates close agreement at the portfolio-level between model output and observed reality during these periods. However, and at the outer fringes of spell life, it is clear that the Cox-models produce outputs that exceed the values of the actual term-structures, regardless of technique. Such overprediction does indeed exaggerate the U-shape in $f_\mathrm{P}$ across $t$ vs that of $f_\mathrm{A}$, which detracts from overall model accuracy. However, this overprediction is at least conservative and risk-prudent in that a bank would prefer greater estimates over lower ones in providing adequately for credit risk.
As measured by the MAE, the degree of the average discrepancy between $f_\mathrm{A}(t)$ and $f_\mathrm{P}(t)$ over $t$ does not seem to vary significantly across the TFD-- and PWP-techniques; both of which approximate an MAE-value of 0.05\%. The exception is the AG-technique with its MAE of 0.0803\%, whose predictions are noticeably less accurate than those of the other techniques.
Furthermore, the PWP-technique does appear to produce an expected term-structure with the lowest MAE-value, even if only marginally so relative to the TFD-technique; i.e., 0.0502\% (PWP) vs 0.0526\% (TFD). What little benefit exists in choosing the PWP-technique over the TFD-technique is largely attributed to the former's ability to attune the baseline hazard to subsequent spells. More importantly, the AG-technique clearly underperforms our expectations and should be duly discarded in favour of the other techniques. It would appear that the way in which time is recorded (calendar time vs gap/spell time) matters to prediction accuracy, in addition to assuming a common baseline hazard across subsequent spells.

%% file: 6-Conclusion.tex
\section{Conclusion}
\label{sec:conclusion}

Default survival modelling with recurrent events has not enjoyed much attention in the literature of credit risk modelling.
We therefore contributed an empirically-driven and pedagogical comparative study amongst three Cox-regression modelling techniques in predicting the time to default over multiple spells. Each technique was fit to a data-rich portfolio of residential mortgages from the South African credit market as an illustration. These techniques include the following. Firstly, the \textit{time to first default} (TFD) Cox-model deliberately ignores recurrent default events, which represented our baseline model within the comparative study. Secondly, the \textit{Andersen-Gill} (AG) Cox-model handles recurrent events by encoding the timing of these events/spells using calendar time over loan life. However, the AG-model cannot easily incorporate spell-specific effects since it assumes a common baseline hazard function across all subsequent spells. This assumption is relaxed by the \textit{Prentice-Williams-Peterson} (PWP) Cox-model, which posits a baseline hazard for each spell number whilst encoding time using the counting process style; i.e., the time spent in the performing spell. All three techniques are fit using a diverse set of input variables, including macroeconomic covariates, which inspired a greater understanding of the underlying drivers of default risk.

Our comparative study of Cox-models is accompanied by a novel suite of diagnostics, which forms the basis of our tutorial. One of these diagnostics is a simple statistical tool by which sampling representativeness can be measured within survival data. This self-styled \textit{resolution rate of type} $\psi$, denoted as $r_\psi\left(t',\mathcal{D}'\right)$, can be computed for any survival dataset $\mathcal{D}'$ over calendar time $t'$. In so doing, a time series is created for each resampled dataset, whereafter discrepancies between two such series can be summarised using the \textit{mean absolute error} (MAE). Smaller MAE-values indicate greater representativeness, and vice versa. Our application of the resolution rate has shown that the resampling scheme is indeed representative of the raw data, which we have subsampled into training and validation sets.
Another contribution is the "clustered tROC-extension", which extends time-dependent \textit{receiver operating characteristic} (ROC) analysis. This type of ROC-analysis is one of the primary ways by which a Cox-model's discriminatory power is analysed over certain time horizons. 
Our extension can contend with the fact that observations within our survival data are clustered around a specific spell of a loan, instead of being completely independent from one another.
Lastly, we contributed a simple method by which the term-structure of default risk can be calculated from the loan-level estimates given by the Cox-models. This predicted term-structure may then be compared against the actual term-structure, where the latter is produced using nonparametric Kaplan-Meier survival analysis. The discrepancy between either term-structure is again summarised using the MAE such that smaller values indicate greater model accuracy.
We conclude our selection of diagnostics with measures of goodness-of-fit, i.e., \textit{Akaike Information Criterion} (AIC), and the \textit{Kolmogorov-Smirnov} (KS) test statistic; as well as Harrell's $c$-statistic of discriminatory power.

As to the question of whether recurrent default events matter in Cox-modelling, we have obtained mixed results. On the one hand, the resolution rate differs by spell number, which suggests that the payment experience is indeed different amidst subsequent cycles of curing and re-defaulting. However, this rather intuitive result does not resurface from the modelling results. We found that the three different Cox-models achieve remarkably similar levels of both goodness-of-fit and discriminatory power, even when measuring the latter over different time horizons using our clustered tROC-extension. That said, the PWP-technique slightly outperforms the AG-technique, which implies that subtle differences exist in the baseline hazard function across subsequent spells. We corroborate this finding by comparing the actual vs expected term-structures resulting from each Cox-model. Again, the difference in MAE-values is negligible between the TFD-- and PWP-techniques, with the PWP-technique outperforming the former only marginally. However, the AG-technique underperformed our expectations quite substantially, which implies that assuming a common baseline hazard is a subpar modelling choice.
Given the minute difference in results between the TFD-- and PWP-techniques, we find that there is little benefit to including recurrent default events into Cox-modelling, at least within our particular dataset. However, this finding certainly depends on the prevalence of recurrent defaults, and we note that only a paltry 7\% of loans in our sample have experienced multiple defaults. At the very least, our tutorial demonstrates three different techniques to handling recurrent default events in any credit dataset.

Future researchers can replicate our study on other types of loan portfolios, particularly those that exhibit a greater prevalence of recurrent default events.
Other researchers may dedicate themselves to refining the way in which the expected term-structure is estimated from a Cox-model. Whilst simplistic, our approach of taking the average event probability at each period has its flaws. In particular, the resulting term-structure does not sum to one over all periods, especially so at later spell periods. This implies that the overall set of average event probabilities are not well-behaved and can break the axioms of probability, particularly at the extremities. In contrast, the actual term-structure that derives from a Kaplan-Meier analysis does sum to one over time, by design.
Another avenue of future work may expand our comparative study to include another recurrent event technique: the \textit{Wei-Lin-Weissfeld} (WLW) technique. While \citet{kelly2000survival} caution against the WLW-technique since it ignores the ordering of recurrent phenomena, a direct comparison might still be worthwhile in the interest of scientific inquiry.
The thematic variable selection process that we have followed may also be augmented with screening the Schoenfeld residuals of each covariate. Doing so may lead to even more parsimonious models by checking the proportional hazards assumption, though possibly at the cost of discriminatory power.
Future work can also focus on studying the statistical properties of our clustered tROC-extension.
Overall, we believe that our work enhances the current practice of Cox-modelling in estimating the lifetime term-structures of default risk under IFRS 9.

%% file: 7-Appendix.tex
\section{Appendix}
\label{sec:Appendix}

In \autoref{app:DataStructures}, we discuss and demonstrate the various data structures at play amongst our survival modelling techniques. Thereafter, the selected input variables are described in \autoref{app:InputSpace}, having followed our thematic variable selection process.

\input{7.1-DataStructures}

\input{7.2-InputSpace}

%% file: 7.1-DataStructures.tex
\subsection{Structuring data according to each recurrent event survival model}
\label{app:DataStructures}

Given the time-dependent nature of our survival data, we illustrate across the following tables the expanded data structure respective to each recurrent survival modelling technique, having assumed time-varying covariates. For the\textit{ time to first default event} (TFD) definition, the stop time $\tau_s$ simply records the loan age at which the spell ended, whereas $\tau_e=0$ will always hold, as illustrated in \autoref{tab:dataStructure_perfSpells_TFD}. The exact same setup holds for the first spell $j=1$ in following the \textit{Anderson-Gill} (AG) technique, shown in \autoref{tab:dataStructure_perfSpells_AG}. Thereafter, the entry and stop times $\tau_e'$ and $\tau_s'$ become the calendar time or loan age at which subsequent spells will start and stop respectively. Under the \textit{Prentice-Williams-Peterson} (PWP) technique, the entry time $\tau_e$ resets to zero upon entering each new performing spell, while the stop time $\tau_s$ resolves into the spell age.
E.g., consider loan 3 with its two performing spells. Under the AG-technique, the timing of these two spells would be recorded as $t_{i1}\in(0,4]$ and $t_{i2}\in (10,13]$, whereas the timings become $t_{i1}\in(0,4]$ and $t_{i,2}\in(0,3]$ under the PWP-technique. Only the timings differ between these two techniques, while the actual lengths of time spent within each spell remain unchanged.

\begin{longtable}[ht!]{p{1cm} p{1.1cm} p{1.6cm} p{1.3cm} p{1.2cm} p{1.2cm} p{1.9cm} p{1.3cm}}
\caption{Illustrating the structure of the raw panel dataset $\mathcal{D}$ and its performing spells for the TFD-technique. The alternating grey-shaded rows indicate loan-level history, while the alternating colour-shaded cells signify the performing spell-level histories respective to each loan; the remaining unshaded cells denote period-level information. Loans 3--4 have multiple spells that are truncated in this technique.} \label{tab:dataStructure_perfSpells_TFD} \\
\toprule
\textbf{Loan} $i$ & \textbf{Period} $t_i$ & \textbf{Spell number} $j$ & \textbf{Spell period} $t_{ij}$ & \textbf{Entry time} $\tau_e$ & \textbf{Stop time} $\tau_s$ & \textbf{Resolution type} $\mathcal{R}_{ij}$ & \textbf{Spell age} $T_{ij}$ \\ 
\midrule
\endfirsthead
\caption[]{(continued)} \\
\toprule
\textbf{Loan} $i$ & \textbf{Period} $t_i$ & \textbf{Spell number} $j$ & \textbf{Spell period} $t_{ij}$ & \textbf{Entry time} $\tau_e$ & \textbf{Stop time} $\tau_s$ & \textbf{Resolution type} $\mathcal{R}_{ij}$ & \textbf{Spell age} $T_{ij}$ \\ 
\midrule
\endhead
\midrule \multicolumn{4}{r}{\textit{Continued on next page}} \\
\endfoot
\bottomrule
\endlastfoot
\cellcolor[HTML]{EFEFEF}1 & 1 & \cellcolor[HTML]{ECF4FF}1 & 1 & \cellcolor[HTML]{ECF4FF}0 & \cellcolor[HTML]{ECF4FF}4 & \cellcolor[HTML]{ECF4FF}1: Defaulted & \cellcolor[HTML]{ECF4FF}4 \\
\cellcolor[HTML]{EFEFEF}1 & 2 & \cellcolor[HTML]{ECF4FF}1 & 2 & \cellcolor[HTML]{ECF4FF}0 & \cellcolor[HTML]{ECF4FF}4 & \cellcolor[HTML]{ECF4FF}1: Defaulted & \cellcolor[HTML]{ECF4FF}4 \\
\cellcolor[HTML]{EFEFEF}1 & 3 & \cellcolor[HTML]{ECF4FF}1 & 3 & \cellcolor[HTML]{ECF4FF}0 & \cellcolor[HTML]{ECF4FF}4 & \cellcolor[HTML]{ECF4FF}1: Defaulted & \cellcolor[HTML]{ECF4FF}4 \\
\cellcolor[HTML]{EFEFEF}1 & 4 & \cellcolor[HTML]{ECF4FF}1 & 4 & \cellcolor[HTML]{ECF4FF}0 & \cellcolor[HTML]{ECF4FF}4 & \cellcolor[HTML]{ECF4FF}1: Defaulted & \cellcolor[HTML]{ECF4FF}4 \\
\cellcolor[HTML]{C0C0C0}2 & 1 & \cellcolor[HTML]{C0DAFE}1 & 1 & \cellcolor[HTML]{C0DAFE}0 & \cellcolor[HTML]{C0DAFE}3 & \cellcolor[HTML]{C0DAFE}4: Censored & \cellcolor[HTML]{C0DAFE}3 \\
\cellcolor[HTML]{C0C0C0}2 & 2 & \cellcolor[HTML]{C0DAFE}1 & 2 & \cellcolor[HTML]{C0DAFE}0 & \cellcolor[HTML]{C0DAFE}3 & \cellcolor[HTML]{C0DAFE}4: Censored & \cellcolor[HTML]{C0DAFE}3 \\
\cellcolor[HTML]{C0C0C0}2 & 3 & \cellcolor[HTML]{C0DAFE}1 & 3 & \cellcolor[HTML]{C0DAFE}0 & \cellcolor[HTML]{C0DAFE}3 & \cellcolor[HTML]{C0DAFE}4: Censored & \cellcolor[HTML]{C0DAFE}3 \\
\cellcolor[HTML]{EFEFEF}3 & 1 & \cellcolor[HTML]{ECF4FF}1 & 1 & \cellcolor[HTML]{ECF4FF}0 & \cellcolor[HTML]{ECF4FF}4 & \cellcolor[HTML]{ECF4FF}1: Defaulted & \cellcolor[HTML]{ECF4FF}4 \\
\cellcolor[HTML]{EFEFEF}3 & 2 & \cellcolor[HTML]{ECF4FF}1 & 2 & \cellcolor[HTML]{ECF4FF}0 & \cellcolor[HTML]{ECF4FF}4 & \cellcolor[HTML]{ECF4FF}1: Defaulted & \cellcolor[HTML]{ECF4FF}4 \\
\cellcolor[HTML]{EFEFEF}3 & 3 & \cellcolor[HTML]{ECF4FF}1 & 3 & \cellcolor[HTML]{ECF4FF}0 & \cellcolor[HTML]{ECF4FF}4 & \cellcolor[HTML]{ECF4FF}1: Defaulted & \cellcolor[HTML]{ECF4FF}4 \\
\cellcolor[HTML]{EFEFEF}3 & 4 & \cellcolor[HTML]{ECF4FF}1 & 4 & \cellcolor[HTML]{ECF4FF}0 & \cellcolor[HTML]{ECF4FF}4 & \cellcolor[HTML]{ECF4FF}1: Defaulted & \cellcolor[HTML]{ECF4FF}4 \\
\cellcolor[HTML]{C0C0C0}4 & 5 & \cellcolor[HTML]{C0DAFE}1 & 5 & \cellcolor[HTML]{C0DAFE}4 & \cellcolor[HTML]{C0DAFE}9 & \cellcolor[HTML]{C0DAFE}1: Defaulted & \cellcolor[HTML]{C0DAFE}5 \\
\cellcolor[HTML]{C0C0C0}4 & 6 & \cellcolor[HTML]{C0DAFE}1 & 6 & \cellcolor[HTML]{C0DAFE}4 & \cellcolor[HTML]{C0DAFE}9 & \cellcolor[HTML]{C0DAFE}1: Defaulted & \cellcolor[HTML]{C0DAFE}5 \\
\cellcolor[HTML]{C0C0C0}4 & 7 & \cellcolor[HTML]{C0DAFE}1 & 7 & \cellcolor[HTML]{C0DAFE}4 & \cellcolor[HTML]{C0DAFE}9 & \cellcolor[HTML]{C0DAFE}1: Defaulted & \cellcolor[HTML]{C0DAFE}5 \\
\cellcolor[HTML]{C0C0C0}4 & 8 & \cellcolor[HTML]{C0DAFE}1 & 8 & \cellcolor[HTML]{C0DAFE}4 & \cellcolor[HTML]{C0DAFE}9 & \cellcolor[HTML]{C0DAFE}1: Defaulted & \cellcolor[HTML]{C0DAFE}5 \\
\cellcolor[HTML]{C0C0C0}4 & 9 & \cellcolor[HTML]{C0DAFE}1 & 9 & \cellcolor[HTML]{C0DAFE}4 & \cellcolor[HTML]{C0DAFE}9 & \cellcolor[HTML]{C0DAFE}1: Defaulted & \cellcolor[HTML]{C0DAFE}5 \\
\end{longtable}

\begin{longtable}[ht!]{p{1cm} p{1.1cm} p{1.6cm} p{1.3cm} p{1.2cm} p{1.2cm} p{1.9cm} p{1.3cm}}
\caption{Illustrating the structure of the panel dataset $\mathcal{D}$ and its performing spells for the AG-technique. Table design follows that of \autoref{tab:dataStructure_perfSpells_TFD}.} \label{tab:dataStructure_perfSpells_AG} \\
\toprule
\textbf{Loan} $i$ & \textbf{Period} $t_i$ & \textbf{Spell number} $j$ & \textbf{Spell period} $t_{ij}$ & \textbf{Entry time} $\tau_e'$ & \textbf{Stop time} $\tau_s'$ & \textbf{Resolution type} $\mathcal{R}_{ij}$ & \textbf{Spell age} $T_{ij}$ \\ 
\midrule
\endfirsthead
\caption[]{(continued)} \\
\toprule
\textbf{Loan} $i$ & \textbf{Period} $t_i$ & \textbf{Spell number} $j$ & \textbf{Spell period} $t_{ij}$ & \textbf{Entry time} $\tau_e'$ & \textbf{Stop time} $\tau_s'$ & \textbf{Resolution type} $\mathcal{R}_{ij}$ & \textbf{Spell age} $T_{ij}$ \\ 
\midrule
\endhead
\midrule \multicolumn{8}{r}{\textit{Continued on next page}} \\
\endfoot
\bottomrule
\endlastfoot
\cellcolor[HTML]{EFEFEF}1 & 1 & \cellcolor[HTML]{ECF4FF}1 & 1 & \cellcolor[HTML]{ECF4FF}0 & \cellcolor[HTML]{ECF4FF}4 & \cellcolor[HTML]{ECF4FF}1: Defaulted & \cellcolor[HTML]{ECF4FF}4 \\
\cellcolor[HTML]{EFEFEF}1 & 2 & \cellcolor[HTML]{ECF4FF}1 & 2 & \cellcolor[HTML]{ECF4FF}0 & \cellcolor[HTML]{ECF4FF}4 & \cellcolor[HTML]{ECF4FF}1: Defaulted & \cellcolor[HTML]{ECF4FF}4 \\
\cellcolor[HTML]{EFEFEF}1 & 3 & \cellcolor[HTML]{ECF4FF}1 & 3 & \cellcolor[HTML]{ECF4FF}0 & \cellcolor[HTML]{ECF4FF}4 & \cellcolor[HTML]{ECF4FF}1: Defaulted & \cellcolor[HTML]{ECF4FF}4 \\
\cellcolor[HTML]{EFEFEF}1 & 4 & \cellcolor[HTML]{ECF4FF}1 & 4 & \cellcolor[HTML]{ECF4FF}0 & \cellcolor[HTML]{ECF4FF}4 & \cellcolor[HTML]{ECF4FF}1: Defaulted & \cellcolor[HTML]{ECF4FF}4 \\
\cellcolor[HTML]{C0C0C0}2 & 1 & \cellcolor[HTML]{C0DAFE}1 & 1 & \cellcolor[HTML]{C0DAFE}0 & \cellcolor[HTML]{C0DAFE}3 & \cellcolor[HTML]{C0DAFE}4: Censored & \cellcolor[HTML]{C0DAFE}3 \\
\cellcolor[HTML]{C0C0C0}2 & 2 & \cellcolor[HTML]{C0DAFE}1 & 2 & \cellcolor[HTML]{C0DAFE}0 & \cellcolor[HTML]{C0DAFE}3 & \cellcolor[HTML]{C0DAFE}4: Censored & \cellcolor[HTML]{C0DAFE}3 \\
\cellcolor[HTML]{C0C0C0}2 & 3 & \cellcolor[HTML]{C0DAFE}1 & 3 & \cellcolor[HTML]{C0DAFE}0 & \cellcolor[HTML]{C0DAFE}3 & \cellcolor[HTML]{C0DAFE}4: Censored & \cellcolor[HTML]{C0DAFE}3 \\
\cellcolor[HTML]{EFEFEF}3 & 1 & \cellcolor[HTML]{ECF4FF}1 & 1 & \cellcolor[HTML]{ECF4FF}0 & \cellcolor[HTML]{ECF4FF}4 & \cellcolor[HTML]{ECF4FF}1: Defaulted & \cellcolor[HTML]{ECF4FF}4 \\
\cellcolor[HTML]{EFEFEF}3 & 2 & \cellcolor[HTML]{ECF4FF}1 & 2 & \cellcolor[HTML]{ECF4FF}0 & \cellcolor[HTML]{ECF4FF}4 & \cellcolor[HTML]{ECF4FF}1: Defaulted & \cellcolor[HTML]{ECF4FF}4 \\
\cellcolor[HTML]{EFEFEF}3 & 3 & \cellcolor[HTML]{ECF4FF}1 & 3 & \cellcolor[HTML]{ECF4FF}0 & \cellcolor[HTML]{ECF4FF}4 & \cellcolor[HTML]{ECF4FF}1: Defaulted & \cellcolor[HTML]{ECF4FF}4 \\
\cellcolor[HTML]{EFEFEF}3 & 4 & \cellcolor[HTML]{ECF4FF}1 & 4 & \cellcolor[HTML]{ECF4FF}0 & \cellcolor[HTML]{ECF4FF}4 & \cellcolor[HTML]{ECF4FF}1: Defaulted & \cellcolor[HTML]{ECF4FF}4 \\
\cellcolor[HTML]{EFEFEF}3 & 11 & \cellcolor[HTML]{E6FFE6}2 & 1 & \cellcolor[HTML]{E6FFE6}10 & \cellcolor[HTML]{E6FFE6}13 & \cellcolor[HTML]{E6FFE6}2: Settled & \cellcolor[HTML]{E6FFE6}3 \\
\cellcolor[HTML]{EFEFEF}3 & 12 & \cellcolor[HTML]{E6FFE6}2 & 2 & \cellcolor[HTML]{E6FFE6}10 & \cellcolor[HTML]{E6FFE6}13 & \cellcolor[HTML]{E6FFE6}2: Settled & \cellcolor[HTML]{E6FFE6}3 \\
\cellcolor[HTML]{EFEFEF}3 & 13 & \cellcolor[HTML]{E6FFE6}2 & 3 & \cellcolor[HTML]{E6FFE6}10 & \cellcolor[HTML]{E6FFE6}13 & \cellcolor[HTML]{E6FFE6}2: Settled & \cellcolor[HTML]{E6FFE6}3 \\
\cellcolor[HTML]{C0C0C0}4 & 5 & \cellcolor[HTML]{C0DAFE}1 & 5 & \cellcolor[HTML]{C0DAFE}4 & \cellcolor[HTML]{C0DAFE}9 & \cellcolor[HTML]{C0DAFE}1: Defaulted & \cellcolor[HTML]{C0DAFE}5 \\
\cellcolor[HTML]{C0C0C0}4 & 6 & \cellcolor[HTML]{C0DAFE}1 & 6 & \cellcolor[HTML]{C0DAFE}4 & \cellcolor[HTML]{C0DAFE}9 & \cellcolor[HTML]{C0DAFE}1: Defaulted & \cellcolor[HTML]{C0DAFE}5 \\
\cellcolor[HTML]{C0C0C0}4 & 7 & \cellcolor[HTML]{C0DAFE}1 & 7 & \cellcolor[HTML]{C0DAFE}4 & \cellcolor[HTML]{C0DAFE}9 & \cellcolor[HTML]{C0DAFE}1: Defaulted & \cellcolor[HTML]{C0DAFE}5 \\
\cellcolor[HTML]{C0C0C0}4 & 8 & \cellcolor[HTML]{C0DAFE}1 & 8 & \cellcolor[HTML]{C0DAFE}4 & \cellcolor[HTML]{C0DAFE}9 & \cellcolor[HTML]{C0DAFE}1: Defaulted & \cellcolor[HTML]{C0DAFE}5 \\
\cellcolor[HTML]{C0C0C0}4 & 9 & \cellcolor[HTML]{C0DAFE}1 & 9 & \cellcolor[HTML]{C0DAFE}4 & \cellcolor[HTML]{C0DAFE}9 & \cellcolor[HTML]{C0DAFE}1: Defaulted & \cellcolor[HTML]{C0DAFE}5 \\
\cellcolor[HTML]{C0C0C0}4 & 20 & \cellcolor[HTML]{B5FFB5}2 & 1 & \cellcolor[HTML]{B5FFB5}19 & \cellcolor[HTML]{B5FFB5}23 & \cellcolor[HTML]{B5FFB5}1: Defaulted & \cellcolor[HTML]{B5FFB5}4 \\
\cellcolor[HTML]{C0C0C0}4 & 21 & \cellcolor[HTML]{B5FFB5}2 & 2 & \cellcolor[HTML]{B5FFB5}19 & \cellcolor[HTML]{B5FFB5}23 & \cellcolor[HTML]{B5FFB5}1: Defaulted & \cellcolor[HTML]{B5FFB5}4 \\
\cellcolor[HTML]{C0C0C0}4 & 22 & \cellcolor[HTML]{B5FFB5}2 & 3 & \cellcolor[HTML]{B5FFB5}19 & \cellcolor[HTML]{B5FFB5}23 & \cellcolor[HTML]{B5FFB5}1: Defaulted & \cellcolor[HTML]{B5FFB5}4 \\
\cellcolor[HTML]{C0C0C0}4 & 23 & \cellcolor[HTML]{B5FFB5}2 & 4 & \cellcolor[HTML]{B5FFB5}19 & \cellcolor[HTML]{B5FFB5}23 & \cellcolor[HTML]{B5FFB5}1: Defaulted & \cellcolor[HTML]{B5FFB5}4 \\
\cellcolor[HTML]{C0C0C0}4 & 40 & \cellcolor[HTML]{FFE1BD}3 & 1 & \cellcolor[HTML]{FFE1BD}39 & \cellcolor[HTML]{FFE1BD}41 & \cellcolor[HTML]{FFE1BD}4: Censored & \cellcolor[HTML]{FFE1BD}2 \\
\cellcolor[HTML]{C0C0C0}4 & 41 & \cellcolor[HTML]{FFE1BD}3 & 2 & \cellcolor[HTML]{FFE1BD}39 & \cellcolor[HTML]{FFE1BD}41 & \cellcolor[HTML]{FFE1BD}4: Censored & \cellcolor[HTML]{FFE1BD}2 \\
\end{longtable}

\begin{longtable}[ht!]{p{1cm} p{1.1cm} p{1.6cm} p{1.3cm} p{1.2cm} p{1.2cm} p{1.9cm} p{1.3cm}}
\caption{Illustrating the structure of the panel dataset $\mathcal{D}$ and its performing spells for the PWP-technique. Table design follows that of \autoref{tab:dataStructure_perfSpells_TFD}.} \label{tab:dataStructure_perfSpells_PWP} \\
\toprule
\textbf{Loan} $i$ & \textbf{Period} $t_i$ & \textbf{Spell number} $j$ & \textbf{Spell period} $t_{ij}$ & \textbf{Entry time} $\tau_e$ & \textbf{Stop time} $\tau_s$ & \textbf{Resolution type} $\mathcal{R}_{ij}$ & \textbf{Spell age} $T_{ij}$ \\ 
\midrule
\endfirsthead
\caption[]{(continued)} \\
\toprule
\textbf{Loan} $i$ & \textbf{Period} $t_i$ & \textbf{Spell number} $j$ & \textbf{Spell period} $t_{ij}$ & \textbf{Entry time} $\tau_e$ & \textbf{Stop time} $\tau_s$ & \textbf{Resolution type} $\mathcal{R}_{ij}$ & \textbf{Spell age} $T_{ij}$ \\ 
\midrule
\endhead
\midrule \multicolumn{8}{r}{\textit{Continued on next page}} \\
\endfoot
\bottomrule
\endlastfoot
\cellcolor[HTML]{EFEFEF}1 & 1 & \cellcolor[HTML]{ECF4FF}1 & 1 & \cellcolor[HTML]{ECF4FF}0 & \cellcolor[HTML]{ECF4FF}4 & \cellcolor[HTML]{ECF4FF}1: Defaulted & \cellcolor[HTML]{ECF4FF}4 \\
\cellcolor[HTML]{EFEFEF}1 & 2 & \cellcolor[HTML]{ECF4FF}1 & 2 & \cellcolor[HTML]{ECF4FF}0 & \cellcolor[HTML]{ECF4FF}4 & \cellcolor[HTML]{ECF4FF}1: Defaulted & \cellcolor[HTML]{ECF4FF}4 \\
\cellcolor[HTML]{EFEFEF}1 & 3 & \cellcolor[HTML]{ECF4FF}1 & 3 & \cellcolor[HTML]{ECF4FF}0 & \cellcolor[HTML]{ECF4FF}4 & \cellcolor[HTML]{ECF4FF}1: Defaulted & \cellcolor[HTML]{ECF4FF}4 \\
\cellcolor[HTML]{EFEFEF}1 & 4 & \cellcolor[HTML]{ECF4FF}1 & 4 & \cellcolor[HTML]{ECF4FF}0 & \cellcolor[HTML]{ECF4FF}4 & \cellcolor[HTML]{ECF4FF}1: Defaulted & \cellcolor[HTML]{ECF4FF}4 \\
\cellcolor[HTML]{C0C0C0}2 & 1 & \cellcolor[HTML]{C0DAFE}1 & 1 & \cellcolor[HTML]{C0DAFE}0 & \cellcolor[HTML]{C0DAFE}3 & \cellcolor[HTML]{C0DAFE}4: Censored & \cellcolor[HTML]{C0DAFE}3 \\
\cellcolor[HTML]{C0C0C0}2 & 2 & \cellcolor[HTML]{C0DAFE}1 & 2 & \cellcolor[HTML]{C0DAFE}0 & \cellcolor[HTML]{C0DAFE}3 & \cellcolor[HTML]{C0DAFE}4: Censored & \cellcolor[HTML]{C0DAFE}3 \\
\cellcolor[HTML]{C0C0C0}2 & 3 & \cellcolor[HTML]{C0DAFE}1 & 3 & \cellcolor[HTML]{C0DAFE}0 & \cellcolor[HTML]{C0DAFE}3 & \cellcolor[HTML]{C0DAFE}4: Censored & \cellcolor[HTML]{C0DAFE}3 \\
\cellcolor[HTML]{EFEFEF}3 & 1 & \cellcolor[HTML]{ECF4FF}1 & 1 & \cellcolor[HTML]{ECF4FF}0 & \cellcolor[HTML]{ECF4FF}4 & \cellcolor[HTML]{ECF4FF}1: Defaulted & \cellcolor[HTML]{ECF4FF}4 \\
\cellcolor[HTML]{EFEFEF}3 & 2 & \cellcolor[HTML]{ECF4FF}1 & 2 & \cellcolor[HTML]{ECF4FF}0 & \cellcolor[HTML]{ECF4FF}4 & \cellcolor[HTML]{ECF4FF}1: Defaulted & \cellcolor[HTML]{ECF4FF}4 \\
\cellcolor[HTML]{EFEFEF}3 & 3 & \cellcolor[HTML]{ECF4FF}1 & 3 & \cellcolor[HTML]{ECF4FF}0 & \cellcolor[HTML]{ECF4FF}4 & \cellcolor[HTML]{ECF4FF}1: Defaulted & \cellcolor[HTML]{ECF4FF}4 \\
\cellcolor[HTML]{EFEFEF}3 & 4 & \cellcolor[HTML]{ECF4FF}1 & 4 & \cellcolor[HTML]{ECF4FF}0 & \cellcolor[HTML]{ECF4FF}4 & \cellcolor[HTML]{ECF4FF}1: Defaulted & \cellcolor[HTML]{ECF4FF}4 \\
\cellcolor[HTML]{EFEFEF}3 & 11 & \cellcolor[HTML]{E6FFE6}2 & 1 & \cellcolor[HTML]{E6FFE6}0 & \cellcolor[HTML]{E6FFE6}3 & \cellcolor[HTML]{E6FFE6}2: Settled & \cellcolor[HTML]{E6FFE6}3 \\
\cellcolor[HTML]{EFEFEF}3 & 12 & \cellcolor[HTML]{E6FFE6}2 & 2 & \cellcolor[HTML]{E6FFE6}0 & \cellcolor[HTML]{E6FFE6}3 & \cellcolor[HTML]{E6FFE6}2: Settled & \cellcolor[HTML]{E6FFE6}3 \\
\cellcolor[HTML]{EFEFEF}3 & 13 & \cellcolor[HTML]{E6FFE6}2 & 3 & \cellcolor[HTML]{E6FFE6}0 & \cellcolor[HTML]{E6FFE6}3 & \cellcolor[HTML]{E6FFE6}2: Settled & \cellcolor[HTML]{E6FFE6}3 \\
\cellcolor[HTML]{C0C0C0}4 & 5 & \cellcolor[HTML]{C0DAFE}1 & 5 & \cellcolor[HTML]{C0DAFE}0 & \cellcolor[HTML]{C0DAFE}5 & \cellcolor[HTML]{C0DAFE}1: Defaulted & \cellcolor[HTML]{C0DAFE}5 \\
\cellcolor[HTML]{C0C0C0}4 & 6 & \cellcolor[HTML]{C0DAFE}1 & 6 & \cellcolor[HTML]{C0DAFE}0 & \cellcolor[HTML]{C0DAFE}5 & \cellcolor[HTML]{C0DAFE}1: Defaulted & \cellcolor[HTML]{C0DAFE}5 \\
\cellcolor[HTML]{C0C0C0}4 & 7 & \cellcolor[HTML]{C0DAFE}1 & 7 & \cellcolor[HTML]{C0DAFE}0 & \cellcolor[HTML]{C0DAFE}5 & \cellcolor[HTML]{C0DAFE}1: Defaulted & \cellcolor[HTML]{C0DAFE}5 \\
\cellcolor[HTML]{C0C0C0}4 & 8 & \cellcolor[HTML]{C0DAFE}1 & 8 & \cellcolor[HTML]{C0DAFE}0 & \cellcolor[HTML]{C0DAFE}5 & \cellcolor[HTML]{C0DAFE}1: Defaulted & \cellcolor[HTML]{C0DAFE}5 \\
\cellcolor[HTML]{C0C0C0}4 & 9 & \cellcolor[HTML]{C0DAFE}1 & 9 & \cellcolor[HTML]{C0DAFE}0 & \cellcolor[HTML]{C0DAFE}5 & \cellcolor[HTML]{C0DAFE}1: Defaulted & \cellcolor[HTML]{C0DAFE}5 \\
\cellcolor[HTML]{C0C0C0}4 & 20 & \cellcolor[HTML]{B5FFB5}2 & 1 & \cellcolor[HTML]{B5FFB5}0 & \cellcolor[HTML]{B5FFB5}4 & \cellcolor[HTML]{B5FFB5}1: Defaulted & \cellcolor[HTML]{B5FFB5}4 \\
\cellcolor[HTML]{C0C0C0}4 & 21 & \cellcolor[HTML]{B5FFB5}2 & 2 & \cellcolor[HTML]{B5FFB5}0 & \cellcolor[HTML]{B5FFB5}4 & \cellcolor[HTML]{B5FFB5}1: Defaulted & \cellcolor[HTML]{B5FFB5}4 \\
\cellcolor[HTML]{C0C0C0}4 & 22 & \cellcolor[HTML]{B5FFB5}2 & 3 & \cellcolor[HTML]{B5FFB5}0 & \cellcolor[HTML]{B5FFB5}4 & \cellcolor[HTML]{B5FFB5}1: Defaulted & \cellcolor[HTML]{B5FFB5}4 \\
\cellcolor[HTML]{C0C0C0}4 & 23 & \cellcolor[HTML]{B5FFB5}2 & 4 & \cellcolor[HTML]{B5FFB5}0 & \cellcolor[HTML]{B5FFB5}4 & \cellcolor[HTML]{B5FFB5}1: Defaulted & \cellcolor[HTML]{B5FFB5}4 \\
\cellcolor[HTML]{C0C0C0}4 & 40 & \cellcolor[HTML]{FFE1BD}3 & 1 & \cellcolor[HTML]{FFE1BD}0 & \cellcolor[HTML]{FFE1BD}2 & \cellcolor[HTML]{FFE1BD}4: Censored & \cellcolor[HTML]{FFE1BD}2 \\
\cellcolor[HTML]{C0C0C0}4 & 41 & \cellcolor[HTML]{FFE1BD}3 & 2 & \cellcolor[HTML]{FFE1BD}0 & \cellcolor[HTML]{FFE1BD}2 & \cellcolor[HTML]{FFE1BD}4: Censored & \cellcolor[HTML]{FFE1BD}2 \\
\end{longtable}

%% file: 7.2-InputSpace.tex
\subsection{A description of selected input variables within each recurrent event Cox-model}
\label{app:InputSpace}

In \autoref{tab:featuresDescription}, the selected input variables of the finalised recurrent event Cox-regression models are described. This description includes a mapping between variables and the specific Cox-model (TFD, AG, and PWP), whilst relegating the various coefficient estimates to the codebase maintained by \citet{botha2025recurrencySourcecode}, purely in the interest of brevity.

\begin{longtable}{p{3.7cm} p{9.3cm} p{2cm}}
\caption{The selected input variables mapped across the various recurrent event Cox-regression models. Subscripts $[\mathrm{a}]$ denote loan account-level variables, $[\mathrm{p}]$ are portfolio-level inputs, and $[\mathrm{m}]$ represent macroeconomic covariates.}
\label{tab:featuresDescription} \\
\toprule
\textbf{Variable} & \textbf{Description} & \textbf{Models} \\ 
\midrule
\endfirsthead
\caption[]{(continued)} \\
\toprule
\textbf{Variable} & \textbf{Description} & \textbf{Models} \\ 
\midrule
\endhead
\midrule \multicolumn{3}{r}{\textit{Continued on next page}} \\
\endfoot
\bottomrule
\endlastfoot
\footnotesize{\texttt{AgeToTerm\_Avg}$_{[\mathrm{p}]}$}  & \footnotesize{Mean value across the portfolio of the ratio between a loan's age and its term.} & \footnotesize{AG, PWP} \\
\footnotesize{\texttt{ArrearsDir\_3\_Changed}$_{[\mathrm{a}]}$}  & \footnotesize{Boolean variable indicating whether a change occurred in the trending direction of the arrears balance over 3 months. This direction is obtained qualitatively by comparing the current arrears-level to that of 3 months ago, binned as: 1) increasing; 2) milling; 3) decreasing (reference); and 4) missing.} & \footnotesize{TFD, AG, PWP} \\
\footnotesize{\texttt{Arrears}$_{[\mathrm{a}]}$} & \footnotesize{Amount in arrears.} & \footnotesize{TFD, AG, PWP} \\
\footnotesize{\texttt{BalanceToPrincipal}$_{[\mathrm{a}]}$} & \footnotesize{Outstanding balance divided by the principal (loan amount) of the loan.} &  \footnotesize{AG, PWP}  \\
\footnotesize{\texttt{g0\_Delinq\_Avg}$_{[\mathrm{p}]}$}  & \footnotesize{Non-defaulted average delinquency $g_0$, as measured using the number of payments in arrears; see the $g_0$-measure from \citet{botha2021paper1}.} & \footnotesize{TFD, AG, PWP}\\
\footnotesize{\texttt{g0\_Delinq\_SD\_4}$_{[\mathrm{a}]}$} & \footnotesize{The sample standard deviation of \texttt{g0\_Delinq} over a rolling 4-month window.} &  \footnotesize{TFD, AG, PWP} \\
\footnotesize{\texttt{InterestRate\_Nominal}$_{[\mathrm{a}]}$} & \footnotesize{Nominal interest rate per annum of a loan.} & \footnotesize{TFD, AG, PWP} \\
\footnotesize{\texttt{LoanType}$_{[\mathrm{a}]}$} & \footnotesize{Echelon of credit market: lower (consumer); upper (wealth).} & \footnotesize{AG} \\
\footnotesize{\texttt{M\_DebtToIncome}$_{[\mathrm{m}]}$}  &\footnotesize{Debt-to-Income: Average household debt expressed as a percentage of household income per quarter, interpolated monthly.} & \footnotesize{TFD} \\
\footnotesize{\texttt{M\_DebtToIncome\_9}$_{[\mathrm{m}]}$} &\footnotesize{9-month lagged version of \texttt{M\_DebtToIncome}.} & \footnotesize{AG, PWP} \\
\footnotesize{\texttt{M\_Inflation\_Growth}$_{[\mathrm{m}]}$} & \footnotesize{Year-on-year growth rate in inflation index (CPI) per month.} & \footnotesize{AG} \\
\footnotesize{\texttt{M\_Inflation\_Growth\_6}$_{[\mathrm{m}]}$} & \footnotesize{6-month lagged version of \texttt{M\_Inflation\_Growth}.} & \footnotesize{PWP} \\ 
\footnotesize{\texttt{M\_RealIncome\_Growth}$_{[\mathrm{m}]}$}  & \footnotesize{Year-on-year growth rate in the 4-quarter moving average of real income per quarter, interpolated monthly.} & \footnotesize{TFD} \\
\footnotesize{\texttt{M\_Repo\_Rate\_6}$_{[\mathrm{m}]}$} & \footnotesize{Prevailing repurchase (or policy) rate set by the South African Reserve Bank (SARB), lagged by 6 months.} & \footnotesize{PWP} \\
\footnotesize{\texttt{PayMethod}$_{[\mathrm{a}]}$} & \footnotesize{A categorical variable designating different payment methods: 1) debit order (reference); 2) salary; 3) payroll or cash; and 4) missing.} & \footnotesize{TFD, AG, PWP} \\
\footnotesize{\texttt{Principal}$_{[\mathrm{a}]}$} & \footnotesize{Principal loan amount.} &  \footnotesize{TFD} \\
\footnotesize{\texttt{Prepaid\_Pc}$_{[\mathrm{a}]}$} & \footnotesize{The prepaid or undrawn fraction of the available credit limit.} & \footnotesize{TFD, AG} \\
\footnotesize{\texttt{RollEver\_24}$_{[\mathrm{a}]}$} & \footnotesize{Number of times that loan delinquency increased during the last 24 months, excluding the current time point.} & \footnotesize{TFD, AG, PWP} \\
\footnotesize{\texttt{Spell\_Num\_Total}$_{[\mathrm{a}]}$} & \footnotesize{The current performance spell number, or total number of visits across all spells over loan life.} & \footnotesize{AG, PWP} \\ 
\footnotesize{\texttt{TimeInDelinqState\_1}$_{[\mathrm{a}]}$} & \footnotesize{Duration (in months) of current delinquency `state' (or value) before the $g_0$-measure changes again in \texttt{g0\_Delinq} to another value, lagged by 1 month.} & \footnotesize{AG} \\ 
\end{longtable}